\documentclass[10pt, conference]{IEEEtran}
\usepackage{cite}
\usepackage{amsmath,amssymb,amsfonts}
\usepackage{graphicx}
\usepackage{textcomp}
\usepackage{xcolor}
\def\BibTeX{{\rm B\kern-.05em{\sc i\kern-.025em b}\kern-.08em
    T\kern-.1667em\lower.7ex\hbox{E}\kern-.125emX}}

\def\junk#1{}
\usepackage{algorithm}
\usepackage{algorithmic}
\usepackage{subfig}

\newcommand{\name}{LIBRA}
\usepackage[normalem]{ulem}

\begin{document}

\title{\vspace{-0.7cm}
\name:~An Economical Hybrid Approach for Cloud Applications with Strict SLAs
}
\author{
\IEEEauthorblockN{Ali Raza}
\IEEEauthorblockA{Boston University\\
araza@bu.edu}
\and
\IEEEauthorblockN{Zongshun Zhang}
\IEEEauthorblockA{Boston University\\
zhangzs@bu.edu}
\and
\IEEEauthorblockN{Nabeel Akhtar}
\IEEEauthorblockA{Akamai Technologies Inc. \\
nakhtar@akamai.com} 

\and
\IEEEauthorblockN{Vatche Isahagian}
\IEEEauthorblockA{IBM Research\\
vatchei@ibm.com}  
\and
\IEEEauthorblockN{Ibrahim Matta}
\IEEEauthorblockA{Boston University\\
matta@bu.edu}
}

\maketitle

\thispagestyle{plain}
\pagestyle{plain}

\begin{abstract}
Function-as-a-Service (FaaS) has recently emerged to reduce the deployment cost of running cloud applications compared to Infrastructure-as-a-Service (IaaS). 
FaaS follows a serverless “pay-as-you-go” computing model; it comes at a higher cost per unit of execution time but typically application functions experience lower provisioning time (startup delay).
IaaS requires the provisioning of Virtual Machines, which typically suffer from longer cold-start delays that cause higher  queuing delays and higher request drop rates.
We present  LIBRA,  a balanced (hybrid) approach that leverages both VM-based and serverless resources to efficiently manage cloud resources for the applications. \mbox{LIBRA} closely monitors the application demand and provisions appropriate VM and serverless resources such that the running cost is minimized and Service-Level Agreements are met.
Unlike state of the art, LIBRA not only hides VM cold-start delays, and hence reduces response time, by leveraging serverless, but also directs a low-rate bursty portion of the demand to serverless where it would be less costly than spinning up new VMs. 
We evaluate LIBRA on real traces in a simulated environment
as well as on the AWS commercial cloud.
Our results show that LIBRA outperforms other resource-provisioning policies, including a recent hybrid approach -- LIBRA achieves
more than 85\% reduction in SLA violations and 
up to 53\% cost savings.

\end{abstract}


\section{Introduction}

In recent years, fueled by increased cloud adoption, business demands, and advances in technology ({\em e.g.}\ AI, edge computing,  etc.), public cloud providers expanded their offerings to include a plethora of services. These services cater for a wide spectrum of customers\footnote{We use the terms ``customer" and ``tenant" interchangeably.} from those who want to control their own development stack to those who want to focus on designing and delivering new features without worrying about the underlying infrastructure. All these services come with varying configurations and have different performance and pricing models. Given all the service options available from the cloud providers, choosing the best suitable service to deploy applications is a challenging problem.

Infrastructure-as-a-Service (IaaS) is one of the most common cloud services available. It allows a tenant to lease VMs with particular configurations to deploy a cloud application. A customer can add or remove VM instances in response to  fluctuating demand of the application.
Depending on the VM configuration and cloud provider, it can take  hundreds of seconds to set up an instance of a VM (cold-start). This can have an adverse effect on the performance of a cloud application, particularly in the presence of flash-crowd traffic demand, where the application demand increases quicker than the VM resource provisioning time. 
To deal with this problem, cloud providers offer auto-scaling service. This service monitors resource utilization ({\em e.g.},\  CPU or memory), and  scales out (increase) or in (decrease) VM instances based on user defined thresholds.

Function-as-a-Service (FaaS) \cite{castro2019rise} is a recent popular offering from most cloud providers. In this model, a tenant writes the code of the application (called serverless function) in one of the supported languages, 
and submits the code to a cloud provider along with dependencies and associated execution triggers.
On an invocation, the code is executed in a sandbox environment, 
and the result is returned to the triggering event. In the past few years, FaaS has evolved to the point that most cloud applications\cite{mark, video_processing_ao,video_processing_zhang,sci-workflow, ser-use-cases,baldini2017serverless,yan2016building,fox2017status} can be developed using FaaS, 
or given the application, it can be easily translated into
the FaaS programming model\cite{translation-ser}. 
A value-added of FaaS is that it does not require explicit scaling instructions from a tenant. The serverless platform is responsible for scaling the resources to address fluctuating demand. 

There are two distinct features that set FaaS apart from IaaS: {\em 1) Quick provisioning:} serverless functions are executed in sandbox environments which can be provisioned within a few milliseconds\cite{peeking-behind}. This feature makes FaaS an ideal service to serve flash-crowd traffic demands. {\em 2) Pricing model:} While both FaaS and IaaS follow the spirit of ``pay as you go" pricing model, FaaS promises real ``pay as you go" with no waste of resources \cite{castro2019rise}.
Under the IaaS model, customers are charged for the entire duration a VM is leased independent of the utilization ({\em i.e.}\ whether or not the application is running). Under the FaaS model, customers are only charged for the execution time of the application function. 
While the cost per unit time of execution in FaaS is comparatively higher than IaaS with the same resources, its cost model -- charging the customer for the precise amount of time an application is running -- makes it an ideal choice for low-duty demand cycles, where leasing a VM instance for a longer time with little to no utilization can be expensive.

Considering the performance and pricing models for IaaS and FaaS, in this paper we show that a combination of IaaS and FaaS is ideal to cater to the dynamic demand of a cloud application.
Previous hybrid approaches \cite{spock,feat,mark} have leveraged the quick-provisioning-time feature of FaaS by directing a portion of the demand to FaaS {\em temporarily} while scaling out IaaS based resources ({\em i.e.}\ VMs), which leads to lower service-level agreement (SLA) violations. To the best of our knowledge, no work has investigated the simultaneous use of FaaS to reduce the overall cost by {\em consistently} directing the low-rate bursty portion of the demand to FaaS.

In this paper, we present \name{}, a load balancing approach for cloud applications with strict SLA requirements\footnote{Examples of such applications are machine learning inference models, IoT applications that collect/process data, etc.}. 
Two contributions set \name{} apart from previous hybrid approaches: 
1) \name{} continually monitors the demand for an application and procures resources in IaaS, FaaS, or both to cater to the demand while optimizing the cost and performance of an application. In contrast, previous hybrid approaches employ VMs as their primary resource and only leverage FaaS to hide VM startup delays;  2) We compare \name{} to Spock~\cite{spock} (a recently proposed hybrid approach)  and other resource provisioning policies. Our evaluation of \name{} on both Amazon Web Services (AWS) and a simulated cloud shows that \name{} outperforms these other approaches in lowering the overall cost of a cloud application while reducing SLA violations in the presence of dynamic demand.

Our contributions are summarized as follows:
\begin{itemize}
    \item In Section~\ref{sec:model}, we present an economic model to analyze the cost of using IaaS versus FaaS to serve a given application demand. We derive a ``cost-indifference point" (CIP) that determines an upper bound on the demand rate to direct to FaaS that guarantees a lower cost compared to using IaaS.
    \item We present the architecture of \name{} (Section~\ref{sec:arch}), a hybrid load balancing approach that {\em simultaneously} uses both FaaS and IaaS, motivated by our analysis that a portion of the demand sent to FaaS at a rate lower than CIP can be served in a cost-effective way.
    \item We evaluate \name{} in a simulated cloud environment (Section~\ref{sec:simulation}) and on AWS services (Section~\ref{sec:aws}) with real application demand traces to establish its efficacy. Our results show that \name{} outperforms other resource provisioning policies, including Spock~\cite{spock}, in reducing both cost (48\% compared to FaaS, 53\% compared to VM over-provisioning, and up to 20\% compared to VM auto-scaling and Spock) and SLA violations (more than 85\% compared to auto-scaling and Spock).
\end{itemize}
Section~\ref{sec:related-work} summarizes related work, and
Section~\ref{sec:conc} concludes the paper.

\section{Background \& Motivation}
\label{sec:model}

Despite development and deployment challenges, IaaS and FaaS are popular choices for building and deploying cloud applications. Typically, developers target one of these services for an application deployment, but recent trends have shown that simultaneous use of these services can improve the application's performance \cite{spock,mark}. 
We contend that simultaneously leveraging both services (IaaS and FaaS) not only improves the performance but can also decrease an application's operational cost. In particular, IaaS is more economical for supporting the high-rate steady portion of the load of application requests, while FaaS is more economical for supporting the low-rate bursty portion of that load. 

FaaS provisions serverless functions in as little as 10s of milliseconds \cite{peeking-behind}.
This quick provisioning time of FaaS has been leveraged in the literature to hide VM cold-start delays and hence lowering SLA violations \cite{spock,feat,mark}.
In contrast to these previous approaches, LIBRA consistently monitors the application load, and directs a portion of the application load to FaaS when it is a cheaper alternative to provisioned VMs.

In this section, we present  an economic (cost) model of an application deployed using either FaaS or IaaS cloud service. 
We use this model to derive the ``cost indifference point" (CIP) as a function of the request arrival rate, where the costs of using IaaS or FaaS for an application are equal.
If the application load is below this CIP value, it is more economical to use FaaS. For higher loads, IaaS is more economical.
This analysis motivates the design of our LIBRA approach.

\begin{figure*}[h!]
\vspace{-3.5ex}
\captionsetup{justification=centering}
 \centering
  \subfloat[AWS Lambda Execution Model]{\includegraphics[width=1.7in]{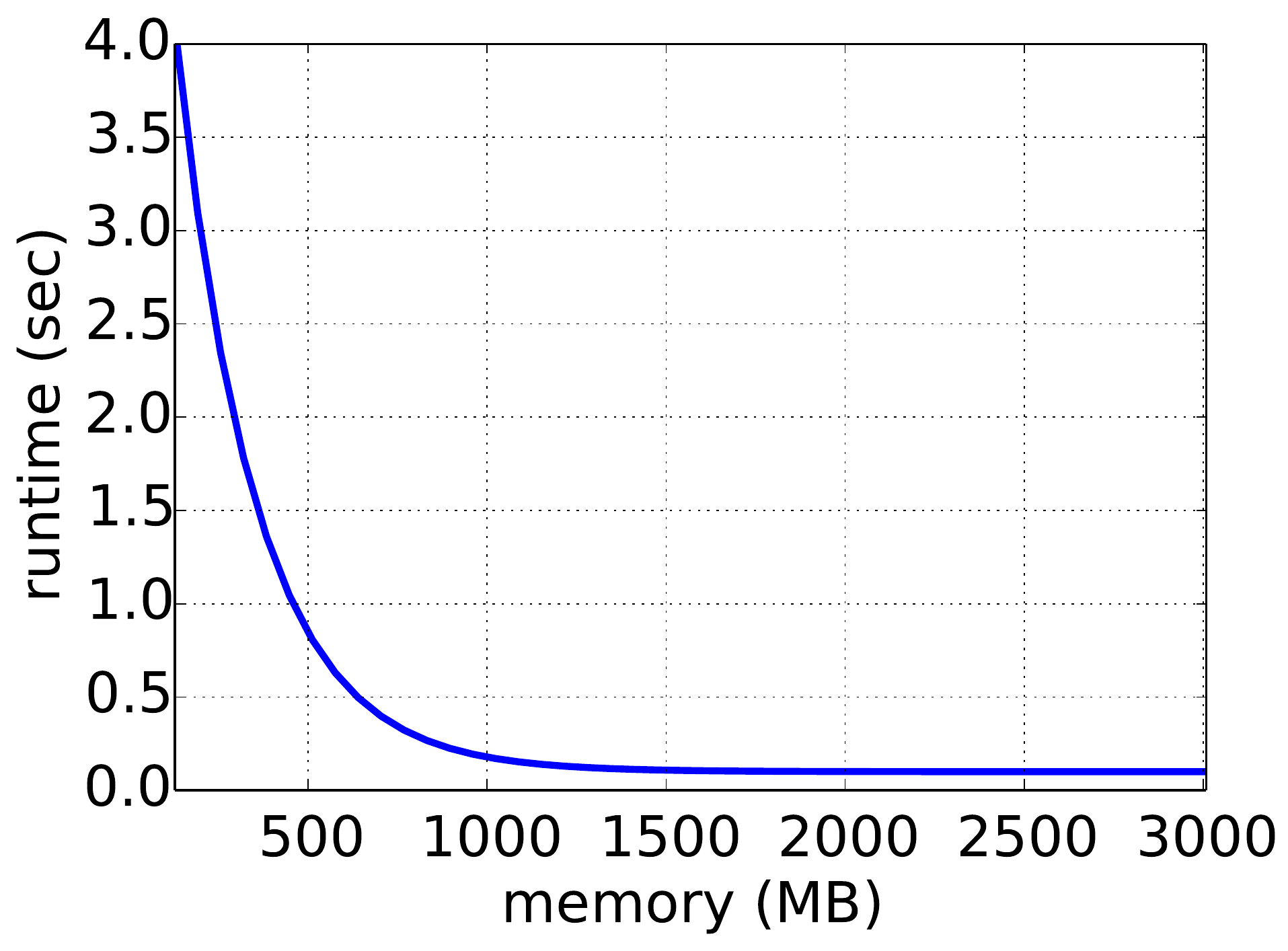}\label{fig:exec_model_fun}  
 \hspace{-0.85em}  }
 \subfloat[Cost at Request Memory=512MB]{\includegraphics[width=1.75in]{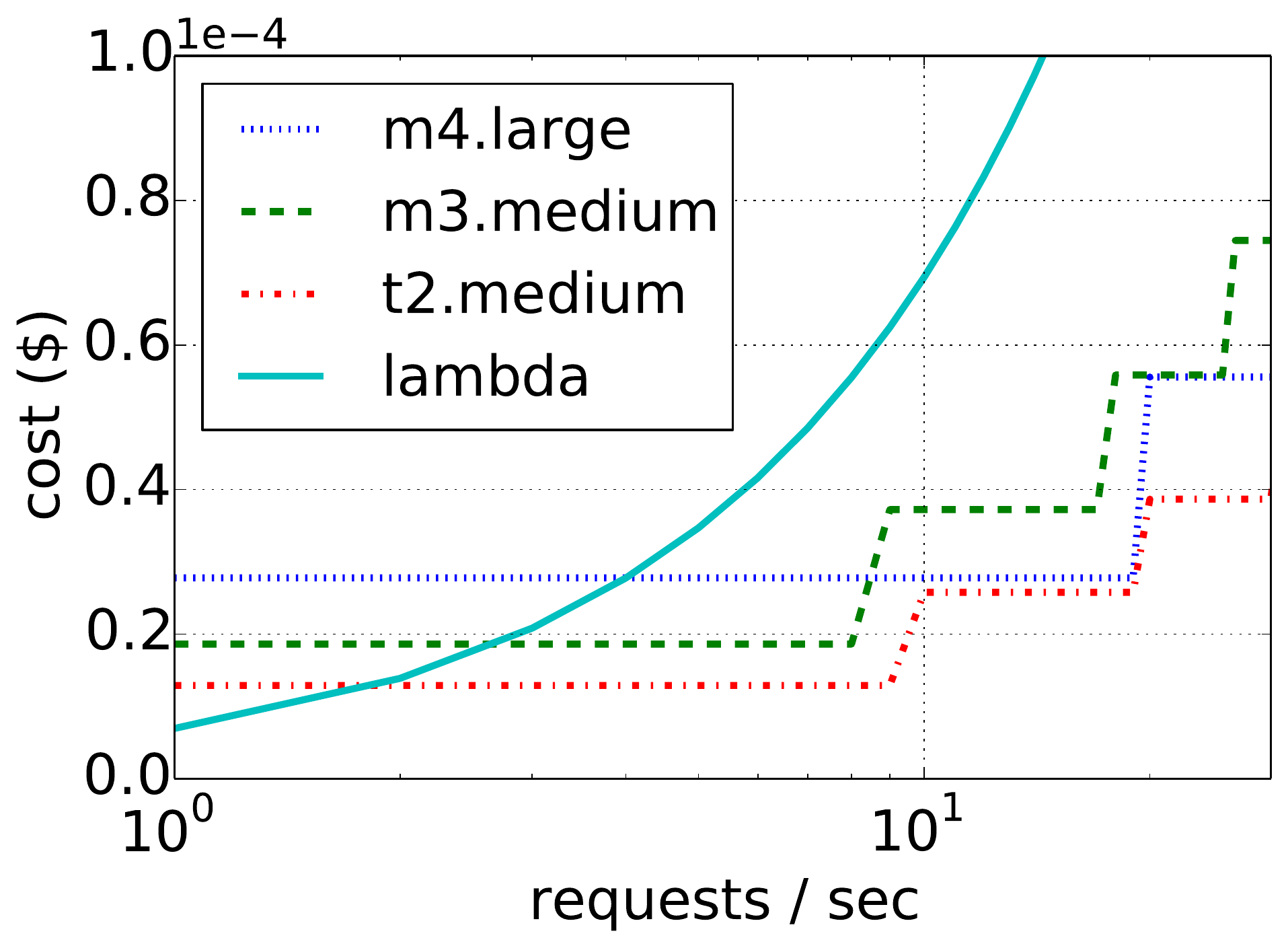}\label{fig:ec2_serverless_512}  
 \hspace{-0.85em}  }
  \subfloat[Cost at Request Memory=3008MB]{\includegraphics[width=1.75in]{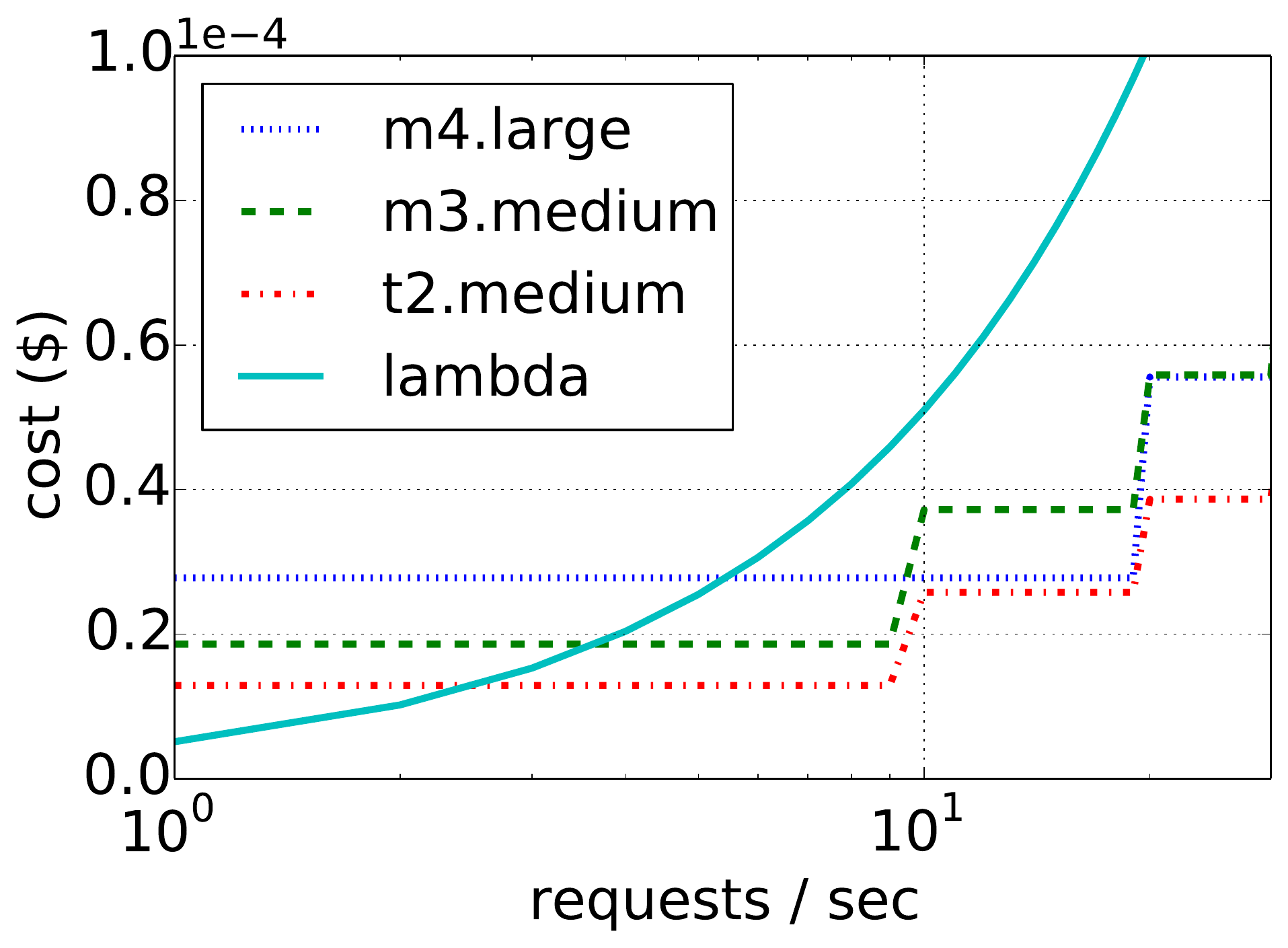}\label{fig:ec2_serverless_1024}  
  \hspace{-0.6em} }
  \subfloat[Hybrid Case]{\includegraphics[width=1.75in]{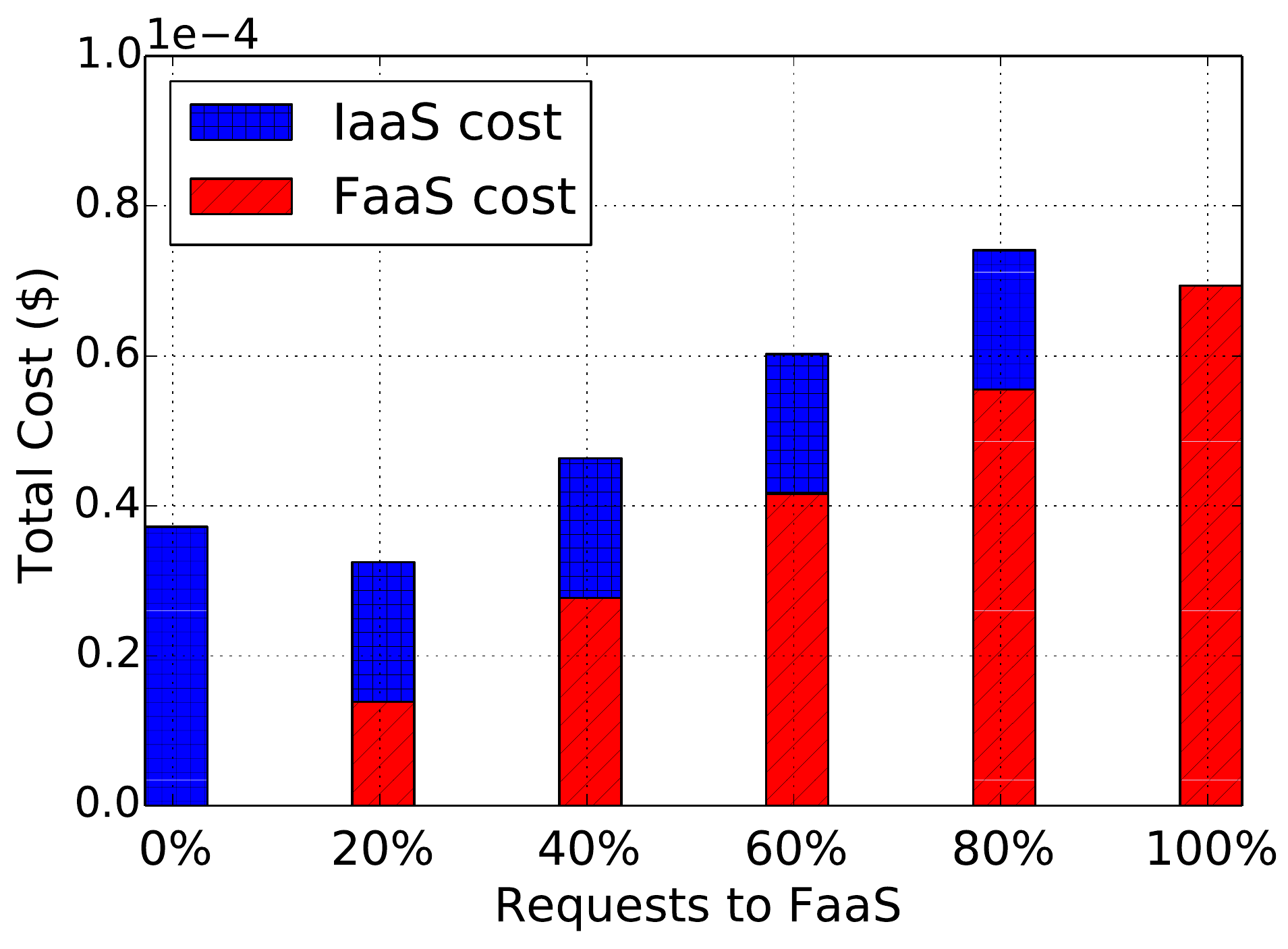}\label{fig:hybrid_motivation}
  \hspace{-0.em}}
\caption{Cost comparison of Amazon Lambda and EC2 instances for varying average request arrival rate} \label{fig:vm_ser_cost_comp}
\vspace{-2.6ex}
\end{figure*}

\subsection{FaaS: Serverless Pricing Model}

Serverless platforms follow a ``pay as you go" pricing model where the user is only charged for the execution time of the serverless function based on a particular configuration ({\em e.g.},\ memory) \cite{aws_lambda_pricing}. 

In~\cite{COSE}, the authors study the effect of configurable resources on various types of serverless functions in AWS Lambda. 
Their results showed that AWS Lambda's execution time follows exponential decay ({\em i.e.},\ diminishing return, as shown in Figure~\ref{fig:exec_model_fun}), and can be expressed as follows:
\begin{align}
t^{FaaS}_{f}(m) & \approx  t^{FaaS}_{f}(m_{max})+ \nonumber\\ 
          &  (t^{FaaS}_{f}(m_{min}) - t^{FaaS}_{f}(m_{max})) \ e^{-\lambda \left(m-m_{min}\right)} \label{exec_model}
\end{align}

\noindent
where $t^{FaaS}_{f}(m)$ is the execution time of a function $f$ when allocated memory $m$ MB, $t^{FaaS}_{f}(m_{min})$ is the running time of $f$ at the smallest possible memory configuration ($m_{min}=128$~MB for Amazon Lambda), $t^{FaaS}_{f}(m_{max})$ is the running time at the largest possible memory configuration ($m_{max}=3008$~MB for Amazon Lambda), and $\lambda$ is a decay constant.

Consider an application, deployed using FaaS, that receives $N$ requests per second,
where each request causes the execution of a serverless function $f$.
The usage cost can be calculated as follows:
\begin{equation}
cost_{FaaS}= \sum_{i=1}^{N}  (t^{FaaS}_{f}(m)\times C_{FaaS}(p, m) + G_{FaaS}(p))
\label{serverless_cost}
\end{equation}
where $C_{FaaS}(p, m)$ is the cost per unit time\footnote{Serverless platforms charge for every 1ms/100ms of execution time\cite{aws_lambda_pricing,ibm_function_pricing, google_function_pricing, azure_function_pricing}.} of executing a serverless function as specified by the serverless platform $p$ for a given configuration $m$,
and $G_{FaaS}(p)$ is the total fixed cost charged by the cloud provider (such as API-gateway cost for AWS Lambda \cite{aws_lambda_pricing}).
This cost model also holds for other cloud providers which follow similar pricing models for FaaS, such as IBM Functions, Google Cloud Functions, {\em etc.}

\subsection{IaaS: VM Pricing Model}

In the IaaS model, a tenant leases a VM with a particular configuration, such as memory, CPU, and storage, to deploy an application. A tenant is charged the cost of the VM independent of its utilization. A VM with a particular configuration can only serve a certain number of requests in a given time period while meeting SLA requirements.

If a VM can host at most $r_{max}$ requests without violating the SLA,
and the IaaS based deployment receives $N$ requests per second, the cost can be calculated as follows: 
\begin{equation}
    cost_{IaaS}= \lceil\frac{N}{r_{max}}\rceil \times C_{vm}(p)
\label{vm_cost}
\end{equation}
where $C_{vm}(p)$ is the cost per unit time\footnote{ IaaS resources can be rented on an hourly basis, while a user can also be charged for partial usage (per second)\cite{azure_compute, amazon_ec2_pricing, ibm_compute}.} of renting a particular VM ($vm$) from a certain cloud provider $p$.

\subsection{Cost Analysis} \label{break-even-point}

Using Equations (\ref{serverless_cost}) and (\ref{vm_cost}),
we compare the cost of deploying an application using
FaaS or IaaS, respectively.
We evaluate the cost for varying demand given by 
$N$, the rate of requests for the application,  where each request causes the invocation of application code deployed using FaaS or IaaS.

The execution model of the application/function used is shown in Figure~\ref{fig:exec_model_fun} and follows an exponential decay ({\em i.e.},\ diminishing return) in the running time of the function with respect to the amount of resources (memory) allocated \cite{COSE},\cite{ishakian2018serving}. It gives the execution time $t_{f}^{FaaS}(m)$ of the application for different memory $m$ settings when deployed using FaaS. IaaS based deployment would follow a slightly different execution model as underlying resources can differ from FaaS.

Thus,
the execution model of the IaaS based deployment with respect to FaaS can be described as:
\begin{equation}
t^{vm}_{f}(m) = \tau \times t^{FaaS}_{f}(m)
\label{exec_model2}
\end{equation}
\noindent
where $t^{vm}_{f}(m)$ is the execution time of the IaaS based deployment when allocated memory $m$ to each request, and $\tau$ is a constant whose value is a real positive number and can vary based on the application and underlying resources. 
Without loss of generality,
we show results where 
under both IaaS and FaaS, the application follows the same execution model ({\em i.e.},\ $\tau=1$), and 
memory is the bottleneck resource in the execution of the function as most FaaS platforms allow only memory as a configurable resource.\footnote{Other resources, such as CPU, I/O, Network, etc., can be bottleneck in the execution of a function. These resources can be substituted here to get similar analysis.}. 
Note that setting $\tau$ to values different than $1$ does not qualitatively affect the results of our analysis.

Using $t^{FaaS}_{f}(m)$ and $m$, 
we calculate $cost_{FaaS}$  
using Equation~(\ref{serverless_cost}), 
where the costs $C_{FaaS}(p, m)$ and $G_{FaaS}(p)$ are taken from AWS Lambda pricing \cite{aws_lambda_pricing}. 

For IaaS, 
the $cost_{IaaS}$ is calculated using Equation~ (\ref{vm_cost}), where $r_{max}$, the maximum number of requests that a VM with memory $M$ can handle in one second, can be derived using Little's Law \cite{allen1990probability}:
\begin{equation*}
    \frac{M}{m} =  r_{max} \times t^{vm}_{f}(m)
\label{little_law}
\end{equation*}
\noindent
$\frac{M}{m}$, the long-term average number of concurrent requests in the system,
equals the arrival rate of these requests ($r_{max}$)
times the (average) time that a request spends in the system ($t^{vm}_{f}(m)$).
We thus have:
\begin{equation}
    r_{max}= \frac{M}{m} \times \frac{1}{t^{vm}_{f}(m)}
\label{max_vm_request}
\end{equation}

We use the AWS Elastic Compute Cloud (EC2) pricing model for different types of EC2 instances (m4.large, m3.medium, and t2.medium). The cost $C_{vm}(p)$ and memory resources $M$ of these instances are specified in EC2 pricing \cite{amazon_ec2_pricing}.
Figure~\ref{fig:ec2_serverless_512} compares the cost of cloud usage when an application is deployed in AWS Lambda or in various instances of EC2 for varying request rate and memory $m$ of 512MB. 
 
We observe that the FaaS model is cost effective when the request rate is below 4 requests/second for the m4.large EC2 instance 
(the point where the m4.large and lambda cost curves intersect). This represents the cost-indifference point (CIP) beyond which the IaaS model is cheaper to be used.
The CIP is obtained by equating Equations~(\ref{serverless_cost}) and (\ref{vm_cost}).
Figures \ref{fig:ec2_serverless_1024} shows a similar behavior when each request is using memory $m$ of 1024MB.

Though the results shown here are obtained using AWS pricing, the cost model is applicable to other cloud services ({\em e.g.},\ from IBM and Google) that follow a similar pricing model. To summarize the key takeaways from our analysis:
\begin{itemize}
    \item The FaaS model is cheaper to use for low duty-cycle application, {\em i.e.}\ when the average request rate $N$ is below the CIP. For higher values of $N$, IaaS is cheaper.
    \item The value of CIP depends on the amount of resources used by each request and the type of VM instance. A tenant can find the appropriate resource configuration by profiling the application or using inference approaches, as proposed in \cite{cherry-pick, COSE}.
\end{itemize}

\subsection{\name{}'s Motivation}

Demand for an application can significantly vary across certain hours of the day and certain days of the week. Based on our analysis in previous sections, an ideal hybrid load balancing approach will have two main characteristics: 
\begin{itemize}
\item[(a)]
It would continually monitor the demand for an application and when the demand is below CIP, it will only provision FaaS resources to cater to the demand as they are more cost-effective in such a scenario. 
This is a feature that previous hybrid approaches ~\cite{spock, mark} lack, as they only use FaaS either for transient demand, or during scaling out VM resources to avoid SLA violations. 

\item[(b)]
It should employ FaaS consistently for a low-rate bursty portion of the demand that the system can not serve using IaaS resources within the SLA. 
This would be beneficial in two ways: 
first, it will reduce the SLA violations, as sudden spikes in demand would be handled by  FaaS, which has negligible cold-start delays and can natively scale-out. 
Second, consistently employing FaaS for a certain portion of the demand can lead to significant cost savings. 
\end{itemize}

To demonstrate the cost saving of such an approach, we leverage the cost analysis in Section~\ref{break-even-point}. 
Consider the scenario shown in Figure~\ref{fig:ec2_serverless_512}, where an application is running on an EC2 instance of the type {\em m3.medium} and it has a steady demand of 10 requests per second where each request requires 512MB of memory. In Figure~\ref{fig:hybrid_motivation}, we compare the cost of serving all the demand or a certain portion of it using FaaS while serving the rest through IaaS. 
We observe that a hybrid approach, where around 20\% requests are served by FaaS and the remaining by IaaS is the most cost-effective as compared to IaaS or FaaS only scenarios. 
This is because $20\%$ of the demand is below the CIP for this particular case and is cheaper to be served through FaaS than 
spinning up a new VM which would be underutilized.

\section{\name~Architecture}
\label{sec:arch}

In this paper, we present  LIBRA,  a  balanced approach that leverages both IaaS and FaaS. It closely monitors the demand from an application and provisions appropriate VM capacity for the IaaS deployment 
to  handle  a  portion  of  the  requests
while directing the rest to be handled by the FaaS based implementation of the application.
Motivated by our analysis in Section~\ref{sec:model}, the design of \name~derives from the following goals:
\begin{itemize}
    \item Serverless functions have quick provisioning time as compared to the IaaS resources where a VM's start-up time can incur 100s of seconds of delay. Thus, it is better to serve transient increase in demand using serverless functions; a FaaS feature exploited in previous approaches \cite{spock,mark,feat}.
    
    \item Always direct a portion of the traffic to serverless functions when it is more cost effective to do so. As explained in Section \ref{break-even-point}, a steady rate of traffic below a certain limit can be cheaper to serve through serverless functions.
\end{itemize}

\begin{figure} [h!]
\begin{center}  
\includegraphics[width=3.2in]{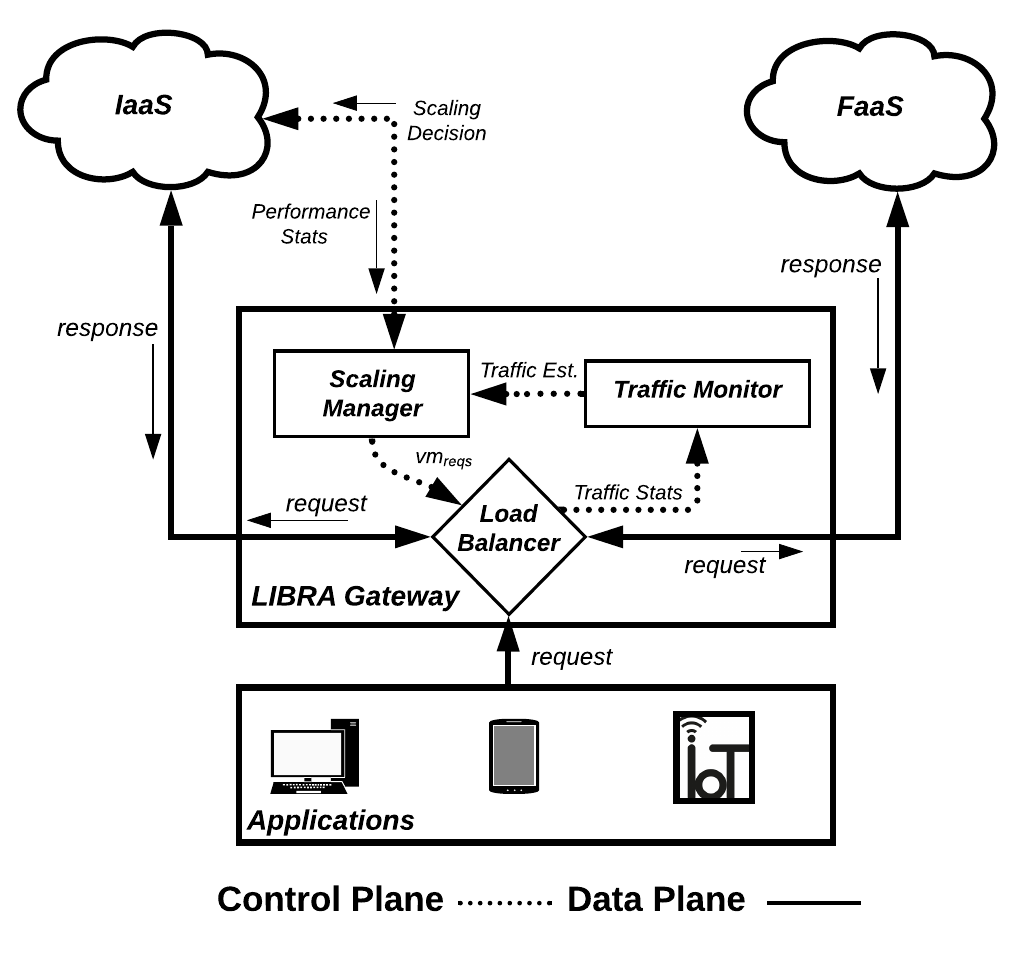} 
\caption{LIBRA architecture}      
\label{fig:arch}
\end{center} 
\vspace{-4ex}
\end{figure}

Figure \ref{fig:arch} gives an overview of our proposed approach. 
The load balancing across  
the IaaS and FaaS based implementations of the application 
is performed through a {\em Load Balancer}, which also updates the traffic statistics and share them with a {\em Traffic Monitor} using the control plane. Based on the traffic demand, a {\em Scaling Manager} provisions VM resources for IaaS 
and updates the Load Balancer to enforce appropriate forwarding rules. Henceforth, we refer to all three components of \name{} as \name{} Gateway (LG). 
\name{} can be deployed by a cloud/service provider as a value-added service or can be directly leveraged by the customers.
In what follows, we explain each component of \name~in detail.

\subsection{Traffic Monitor} \label{traffic-monitor}
The Traffic Monitor (TM) continually receives traffic updates from the {\em Load Balancer} (LB).
Using the historical data of these updates, the TM estimates the future load for the application.
The forehand knowledge of traffic is critical particularly for the VM-based resources, because they can take up to several minutes to start and be ready to serve application traffic. In the current implementation, we introduce the notion of an {\em epoch} that represents a configurable unit-time  ({\em e.g.},\ 10 seconds or 1 minute). The LB continually reports the number of requests received in an {\em epoch} to the TM, which uses this information to 
estimate the future load. 
Currently, the TM keeps track of the Exponentially Weighted Moving Average (EWMA) and sample deviation of requests received in previous epochs as given in Equations (\ref{moving_ave}) and (\ref{moving_std}), where $reqs_{curr}$ is the number of requests received in the current epoch,  $\alpha$ and $\beta$ $\in [0,1]$ are configurable based on how quickly a user wants the system to react in the face of traffic variations. Our experiments have shown that EWMA and sample deviation of the number of requests track well the traffic variation\footnote{ A \name{} user can employ other traffic prediction models as well.} as shown in Figure \ref{fig:traces}. 
\begin{equation}
    avg = (1 -\alpha) \times avg + \alpha \times reqs_{curr} 
\label{moving_ave}
\end{equation}
\begin{equation}
std =  (1 -\beta) \times std + \beta \times \left|reqs_{curr} - avg\right|
\label{moving_std}
\end{equation}
The TM reports the $avg$ and $std$ values to the {\em Scaling Manager} (SM) every $K$ epochs. The SM then makes scaling decisions as explained next. 

\subsection{Scaling Manager}
\label{scaling-manager}

The SM is a crucial component of our \name~architecture. It periodically receives the traffic statistics from the TM and orchestrates resources in the form of VMs and serverless functions to serve application requests.

As one of the design goals of \name~is to keep the serverless load below a certain threshold (CIP) to avoid overpaying, and send the maximum stable load to provisioned VMs for their cost effectiveness, our SM provisions VM (IaaS) resources that can handle a request rate equal to $(avg+ \phi \cdot std)$, with remaining requests directed to run as serverless functions (FaaS). $\phi \in \mathbb{R} $ is a configurable parameter of the \name{} system {\em and is discussed later in Section~\ref{parameters}.}  Our experiments (cf.\ Sections \ref{sec:simulation} and \ref{sec:aws}) have shown that any traffic above $(avg+\phi \cdot std)$ is either transient or cheaper to be served by serverless functions. \name{} provisions VMs cautiously based on the estimated demand given by $(avg+\phi \cdot std)$ so as to avoid either under-provisioning VMs and then suffering from higher startup-delays while spinning up additional VMs, or over-provisioning VMs and paying unnecessary cost due to VM under-utilization.

\begin{algorithm}[h!]
\caption{LIBRA's Scaling Algorithm}\label{scaling-algo}
\textbf{Input:}\\
$avg$, $std$: EWMA and sample deviation reported by the TM \\
$VM_{res}$: resources available in each VM instance\\
$req_{res}$: resources required to serve one application request \\
$req_{time}$: average request service time \\
$cip$: Cost Indifference Point  \\
$\phi$: number of sample deviations beyond average demand

\textbf{Output:} \\
$vm_{reqs}$ // request rate that provisioned VMs can handle  \\
\begin{algorithmic}[1]

\STATE $active\_vms$ = $get\_active\_vms()$ \\
~~~~~~~~~~~~~~~~~~// returns the number of active VMs 
\IF {$avg < cip$}
    \STATE $remove\_vms$($active\_vms$) \\
    ~~~~~~~~~~~~~~// removes all VM instances \\
    $vm_{reqs}$ = 0
    \RETURN
\ENDIF
\STATE $r_{max}$ = $vm\_capacity$($VM_{res}$, $req_{res}$, $req_{time}$) \\
~~~~~~// get maximum number of requests a VM can serve
\STATE $vm_{reqs}$ = $avg + \phi \cdot std$
\STATE $num\_instances$ = $\lceil(vm_{reqs} / r_{max})\rceil$
\STATE $vm\_diff$ = $num\_instances$ - $active\_vms$
\IF {$vm\_diff$ $>$ 0}
    \STATE $add\_vms$($vm\_diff$) // adds VM instances 
\ELSE
    \STATE $remove\_vms$($vm\_diff$) // removes VM instances
\ENDIF

\end{algorithmic}
\label{alg-1}
\end{algorithm}

Algorithm~\ref{alg-1} describes how the SM procures VM resources for IaaS.
In line 1, the function $get\_active\_vms()$ returns the current number of active VMs. 
Line 2 checks if the average request rate is below the CIP threshold (obtained through cost analysis). 
If it is, \name~shuts down and deallocates all the currently provisioned VMs (line 3), as the current demand can be met cost-effectively using only serverless functions. 
In line 6, the algorithm calculates the number of requests ($r_{max}$) that a VM with given resources can serve while meeting the SLA. 
In the case of homogeneous VM resources and consistent workload for each request, $r_{max}$ can be calculated by using Equation~(\ref{max_vm_request}). 
In the case of variable workload,
adaptive controllers (e.g., PID \cite{control-book}) can be used to set $r_{max}$. 
A proportional–integral–derivative (PID) controller can adapt $r_{max}$ based on the error between the target and measured response/service time. In lines 7-8, we obtain the number of VM instances that are needed to cater to a demand $vm_{reqs}= avg+\phi \cdot std$, as instantaneous requests beyond that value are considered transient and will be handled by serverless functions.  
The functions  $add\_vms$ and  $remove\_vms$ (lines 11 and 13) implement the VM-cloud (IaaS) interface to allocate or deallocate VMs to achieve the desired $num\_instances$ (line 8). Moreover, if the decision is to add more VMs, 
the {\em Scaling Manager} waits until the VMs are in ready state before sending a $vm_{reqs}$ update to the {\em Load Balancer}.

\subsection{Load Balancer}
\label{load-balancer}

The LB receives requests from the end-users and forwards them to the appropriate resources, either VMs (IaaS) or serverless (FaaS). It also keeps track of the requests received in an epoch and periodically notifies the {\em Traffic Monitor}. Moreover, whenever the {\em Scaling Manager} makes a scaling decision, it reports the new value of $vm_{reqs}$ to the {\em Load Balancer} as the SM provisions VMs to accommodate a request rate of $vm_{reqs}$.
From queuing theory \cite{qeuing-model}, to ensure stable (predictable) performance and small queuing delays, the request rate to the provisioned VMs should be lower than the service rate given by the VM provisioned rate of $vm_{reqs}$. 
This keeps the aggregate utilization of the
provisioned VMs below one. 
Consequently, our LB directs only a fraction $\rho$ of the request rate, {\em i.e.},\ $\rho \cdot vm_{reqs}$ to the VM resources. {\em This fraction $\rho$ of provisioned VM capacity that can be used to serve requests is a configurable parameter of \name{} and discussed in detail in Section \ref{parameters}}.

The LB adopts a forwarding approach that directs requests to VMs (IaaS) first, which has two key benefits: 1) The VMs are already in ready state and will not incur any cold-start delays, and 2) ready VMs are cheaper compared to serverless.

\subsection{\name{} Parameters} 
\label{parameters}

Our \name{} approach has the following configurable parameters that an administrator can tweak to maximize their gain whether it is performance, cost, or both. We studied the behavior of these parameters in simulation
and experimentally,
and here we briefly summarize the effect of the following parameters and their recommended settings.
\subsubsection{EWMA Weights}
The TM in \name~uses EWMA to monitor the average rate of requests and sample deviation. The weights $\alpha$ and $\beta$ given to the most recent number of requests observed over the current epoch are configurable parameters. A high weight value can lead to a quick response to a sudden increase in demand, resulting in over-provisioning of VM resources if the increase were transient. On the other hand, a low weight value can lead to a slow response to a sudden increase in demand, resulting in under-provisioning of VM resources if the decrease were persistent, which increases the usage of serverless functions and results in a higher cost. 
\subsubsection{Scaling Decision Interval}
The SM makes scaling decisions every $K$ epochs. A smaller value of $K$ ({\em e.g.},\ less than the startup delay of VMs) can result in back-to-back scaling decisions without waiting for the system to react and reach equilibrium.  A large value of $K$ results in longer intervals between scaling decisions, hence slower adaptation leading to missing potential cost savings. The scaling decision interval should be larger than the startup delay of the VM instances being used. This is because when the {\em Scaling Manager} makes the decision to scale out, it waits for the newly added VM instances to be in ready state before informing the {\em Load Balancer} of the newly provisioned VM capacity. 
In our experiments we set the scaling decision interval to be three times the VM startup delay.

\subsubsection{VM Utilization} \label{vm-util}
As described in Section \ref{load-balancer}, the {\em Load Balancer} only utilizes a certain proportion ($\rho$) of the provisioned VMs. The goal is to make sure that the VMs serve the requests without SLA violations
(cf. Section~\ref{load-balancer}). In our experiments we set this parameter $\rho$ to 80\%.

\subsubsection{Traffic Estimation}
To estimate traffic demand, \name{} uses the EWMA and sample deviation to obtain $(avg+ \phi \cdot std)$, where $\phi$ is a real number and its value depends on the fluctuations in demand. In \name{}, IaaS resources are provisioned for $(avg+ \phi \cdot std)$ demand. 
Hence, a higher value of $\phi$ will cause more aggressive provisioning of VMs, which can potentially lead to VM under-utilization. 
On the other hand, a lower value of $\phi$ can lead to more FaaS usage which can potentially lead to higher cost as serving requests by FaaS is expensive.

In the next two sections, 
we evaluate \name~using simulations and on Amazon AWS.

\section{Simulation Model \& Results} 
\label{sec:simulation}

\name{} closely monitors the demand for an application and consequently provisions VM resources, while the transient spikes and small portion of the demand are served by serverless functions. This approach results in little to no SLA violations while also reduces the cost of cloud usage for the tenant. To evaluate the long-term efficacy of \name{}, we modeled~\footnote{The simulator code is available at~\cite{libra_sim}.} various cloud services after Amazon Web Services (AWS) \cite{aws}. These include IaaS, FaaS, Load Balancer, and Autoscaler. Using real traces, we evaluated \name{} against different approaches: VM over-provisioning, FaaS-only, and provisioning of VM resources using the autoscaler. 

\subsection{Modeling Cloud Services}
We modeled IaaS and FaaS (and related services) after  AWS EC2 and Amazon Lambda, respectively. 

\subsubsection{IaaS}
Our modeled IaaS has various resource types to offer for application deployment. Different VM instance types have different cold-start delay depending on the size of the instance and resource such as memory.  
Moreover, our pricing model follows the AWS EC2 pricing model, where users are charged based on partial usage, {\em i.e.}\ on seconds basis as specified in \cite{amazon_ec2_pricing}. 
Any instance can host a pre-defined number of requests based on the resources available in the instance and desired SLA. Hosting more requests on an instance can lead to performance degradation and potential SLA violations.
The usage cost is calculated according to Equation~(\ref{vm_cost}).\\
\indent {\em Load Balancer:} Production-ready applications typically use more than one VM. Incoming requests are distributed among them in a Round Robin fashion. 
If all the VM instances already have a pre-defined number of requests running, any subsequent request is queued and served as soon as any instance can accommodate it. \\
\indent {\em Autoscaler:} Our modeled autoscaler works similarly to Amazon EC2 Auto Scaling \cite{amzone_auto_scaling} and allows users to define auto-scaling policies such as scale-in/scale-out thresholds, scaling groups, and minimum/maximum number of instances. Moreover, our autoscaler can use a threshold on metrics, such as average memory utilization or request count on each instance, to make scaling decisions.
\subsubsection{FaaS}
To model FaaS, we deployed various types of application function on Amazon Lambda and found the relationship between the configurable resources ({\em e.g.},\ memory) and execution time. This follows an exponentially decay model as given by Equation~(\ref{exec_model}). Other approaches ({\em e.g.},\ \cite{COSE},\cite{ishakian2018serving}) have reported similar execution patterns. We also use Amazon Lambda's pricing model where, based on the configured resources and execution time, usage cost is calculated according to Equation~(\ref{serverless_cost}).
\subsubsection{Simulation Parameters}
For our evaluation, we chose the ``large" EC2 instance type {\em m4.large}, which has 8.0 GB of memory and 0.1 dollars per hour cost. 
We ran multiple instances of type {\em m4.large} and noted that the provisioning time (cold-start delay) to obtain an instance is about 100 seconds. 
Each request requires 512MB of memory to complete in one second on both IaaS and FaaS.  For \name{} parameters, we used 0.2 as the EWMA weight, 300 seconds as the scaling decision interval (which is $3\times$ the cold-start of the VM instance being used), VM utilization threshold $\rho$ = 80\%, and 
traffic estimation parameter $\phi = 1$.

\subsection{Log Traces}
\label{sec:sim-trace}
\begin{figure} [h!]
\begin{center}  
\includegraphics[width=3.2in, height=2.3in]{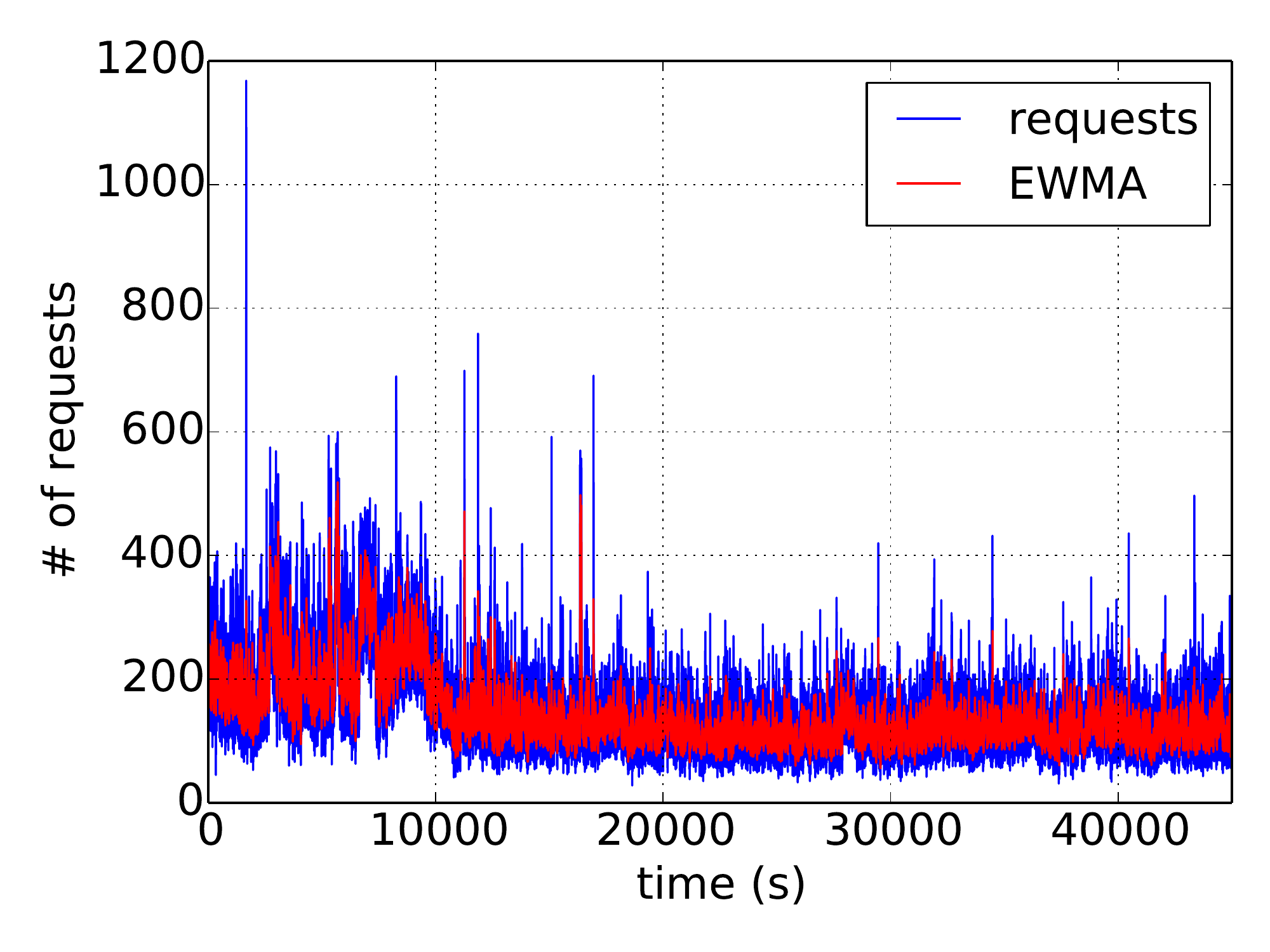} 
\caption{WITS trace and EWMA}   
\label{fig:traces}
\end{center}
\vspace{-4ex}
\end{figure}

We used WITS \cite{wits_trace}, Berkeley \cite{berkley_trace}, and synthetic traces to evaluate the efficacy of \name{}. These traces have also been used to evaluate similar approaches \cite{spock}. We utilize a 12-hour long segment from the traces to generate the workload for our simulation. Each request is assumed to have a constant and equal service (execution) time 
when served on either the IaaS or FaaS based implementation. Our simulator takes into account the cold-start of VMs under IaaS. 
We also assume that the serverless functions under FaaS have a minimal cold-start delay that would not affect the performance of an application with relatively high popularity as shown in \cite{peeking-behind}. While \name{} produced similar results for all traces, due to space limitation, we only present results for the WITS trace. 
Figure \ref{fig:traces} shows the 12-hour snippet of a WITS trace of the number of requests per epoch (second), along with the EWMA. Despite the high variation of the demand, EWMA accurately tracks the dynamicity of the trace.

\subsection{Resource Provisioning \& Deployment Policies}
\label{sec:policies}

As discussed in Section~\ref{sec:arch},
\name's main goal is to utilize both 
FaaS and IaaS to minimize the overall cost while at the same time meeting the performance (execution time) requirement of the application. 
\name{}
leverages the best of both cloud services: 
the quick provisioning time of serverless functions and 
the low cost of provisioned VM resources.
We compare \name's balanced approach to the following policies: 

\subsubsection{Over-provisioning (MAX)}
Cloud applications have strict SLAs and not meeting their performance constraints can result in bad user experience and potentially loss of revenue. To avoid potential SLA violations, the tenant could opt to over-provision the VM resources. We simulate this scenario by scanning the whole demand trace and provisioning the VM resources based on the {\em maximum} number of requests received during a second ({\em i.e.},\ the peak rate). While such approach would avoid SLA violations, allocated VM resources will be underutilized and the client would incur higher costs.
\subsubsection{Autoscaling (AUTO)}
 
Autoscaling is a popular service provided by major cloud providers. In Autoscaling, the performance of the currently allocated resources ({\em e.g.},\ VMs), is monitored based on some metric. The performance metrics can vary based on the cloud provider, but some examples include memory/CPU utilization or request/connection count on each host. If the metric exceeds a certain threshold, new resources are added to the system to avoid potential overloading of current resources and subsequent degradation of performance. If the metric falls below a certain threshold, resources are removed to avoid under-utilization. 

\subsubsection{Spock}

Previous approaches, such as Spock \cite{spock} and MArk~\cite{mark} reduce SLA violations due to VM start-up delays by directing demand to serverless functions while VMs are being provisioned. Unlike LIBRA, Spock-like schemes do {\em not} consistently and simultaneously use serverless functions to serve transient demand and reduce overall cost. We simulated this by directing the {\em excess} portion of the demand to FaaS {\em during} scale-out events.

\subsubsection{FaaS only} 
The application is deployed on a serverless platform and all requests are served by serverless functions. 

\subsection{Simulation Results}
\subsubsection{Cost and SLA violations}

\begin{figure*}[h!]
\vspace{-3.5ex}
\captionsetup{justification=centering}
 \centering

  \subfloat[\name{} and other resource provisioning policies]{\includegraphics[width=1.75in]{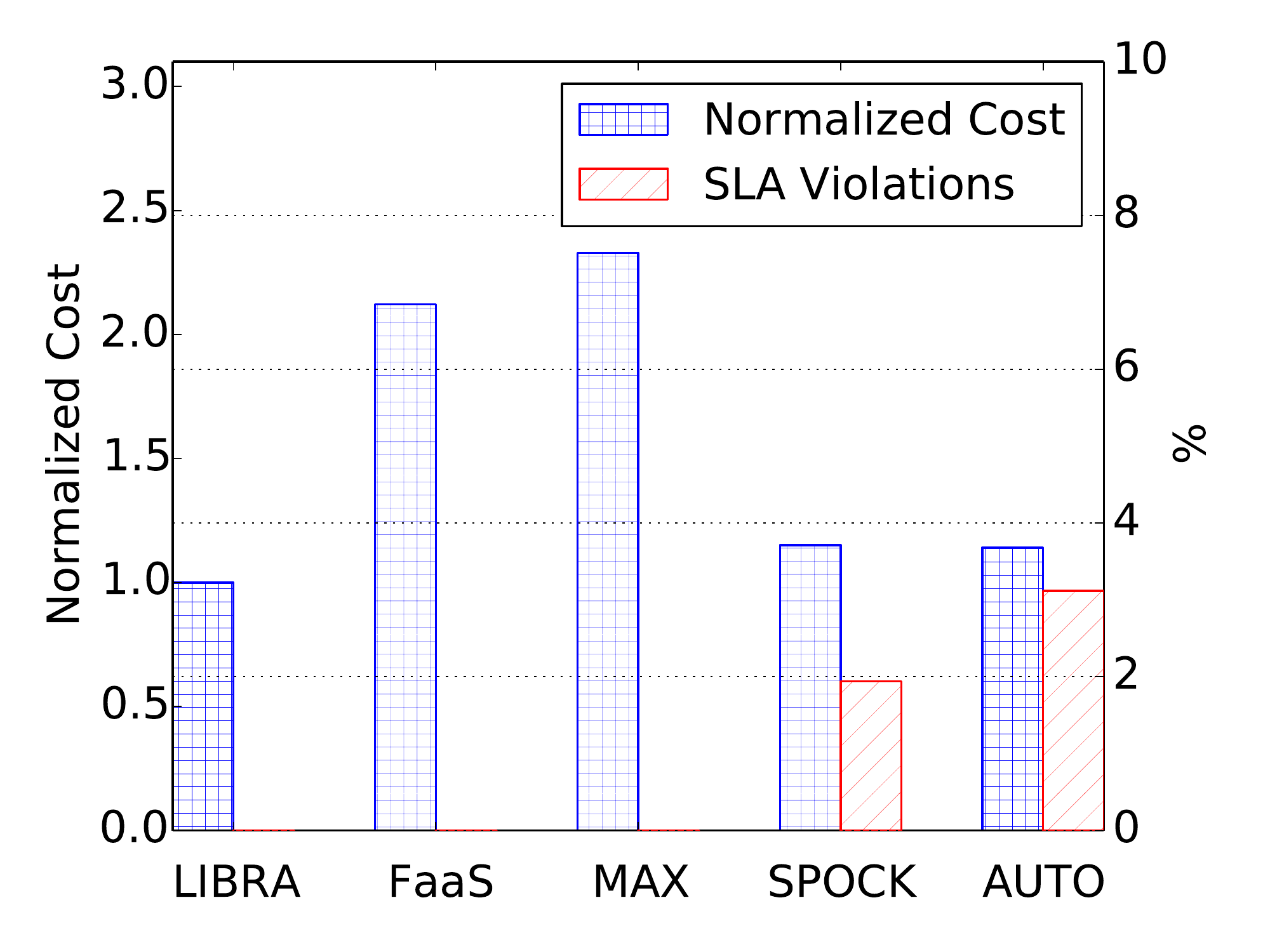}\label{fig:cost_perf}  
 \hspace{-0.85em}  }
 \subfloat[\name{}'s VM provisioning]{\includegraphics[width=1.75in]{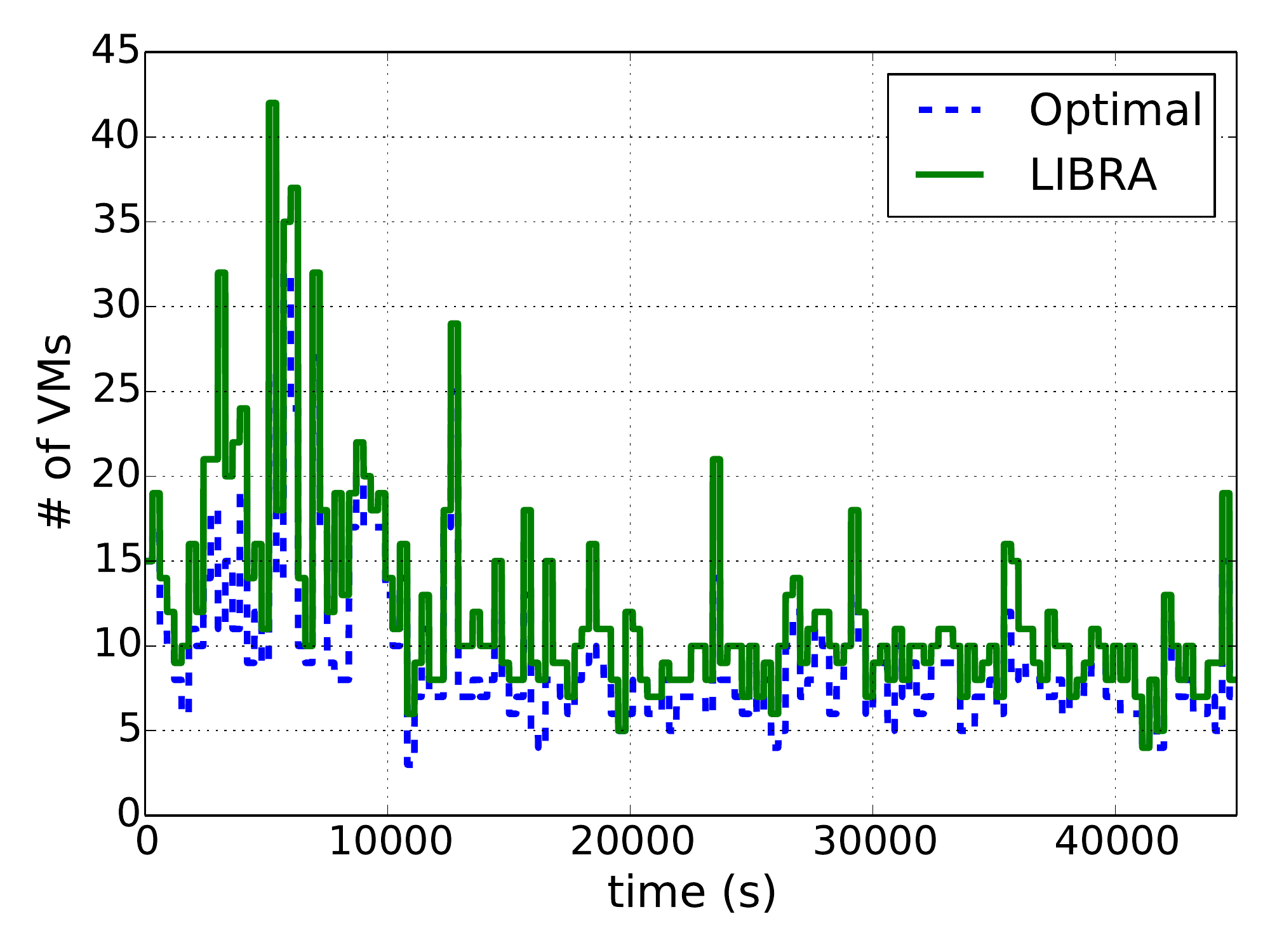}\label{fig:vm_prov}  
 \hspace{-0.85em}  }
  \subfloat[\name{}'s request distribution among available resources]{\includegraphics[width=1.75in]{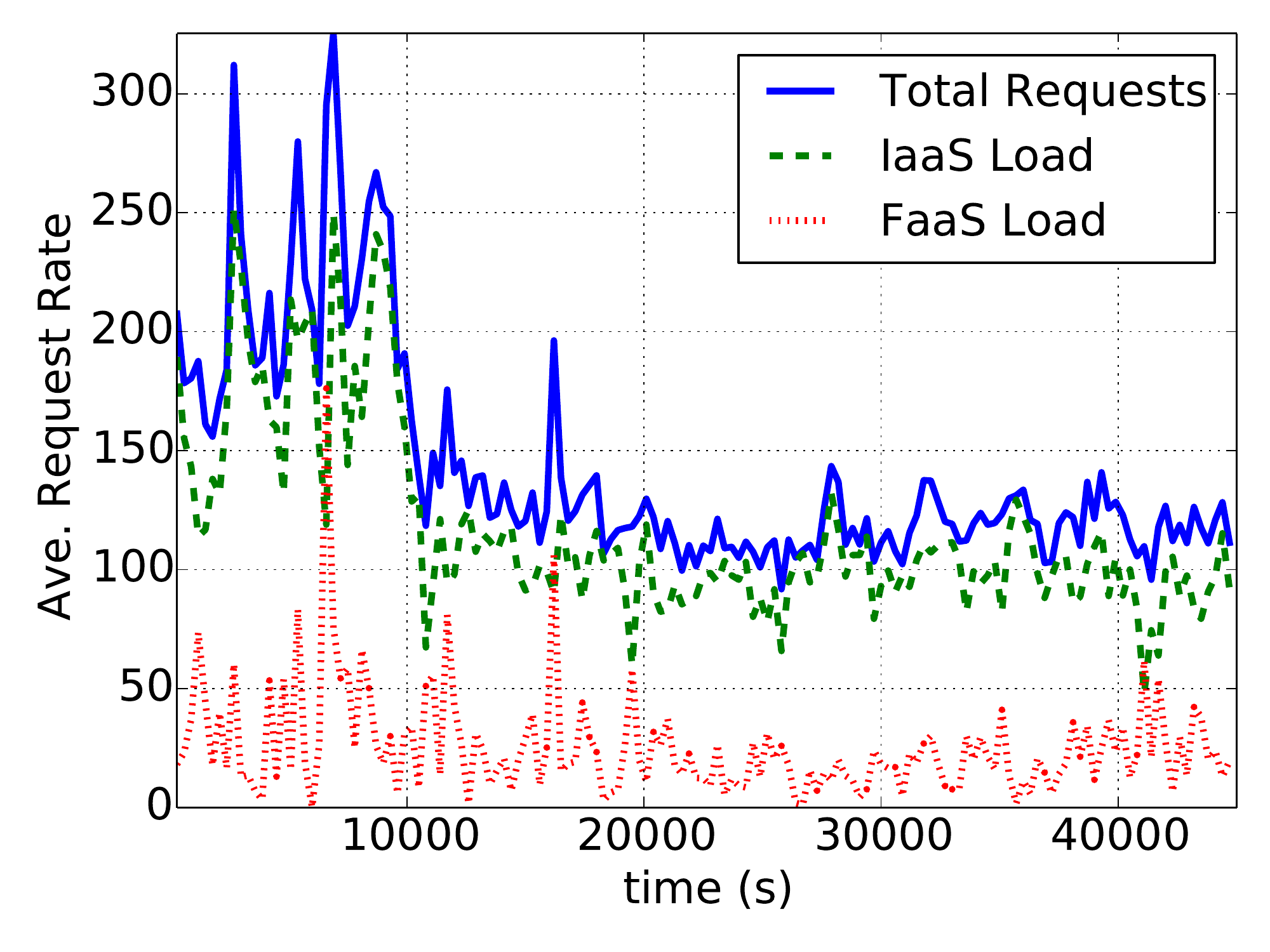}\label{fig:req_dist}  
  \hspace{-0.6em} }
\caption{Cost and performance of \name{} in simulated cloud environment} \label{fig:libra}
\vspace{-2.6ex}
\end{figure*}

Figure \ref{fig:cost_perf} compares the cost and performance of the aforementioned resource provisioning policies with \name{}. The x-axis represents the different approaches described above, the y-axis (left) represents the incurred cost normalized to that of LIBRA, and the y-axis (right) reports the percentage of SLA violations. 
\name{}'s cost also includes the cost of the VM instance used to deploy the \name{} Gateway (cf.\ Figure~\ref{fig:arch}). As expected, over-provisioning (MAX) leads to zero SLA violations, but incurs the highest cost. 
The autoscaling approach (AUTO) reduces cost significantly but introduces significant SLA violations. This is due to the fact that VMs have high cold-start delays, and while a new VM is being set up to share the demand, existing VM instances get saturated, which leads to performance degradation and SLA violations. 
We note that we have experimented with various thresholds for the utilization metrics used by autoscaling. However, we have observed that either the system performs better at the expense of a much higher cost or the system has lower cost at the expense of a much worse performance (higher SLA violations).  
Compared to autoscaling,
Spock reduces SLA violations by more than 35\% at about the same cost. 
Notice that consistent with the original study~\cite{spock}, Spock reduces SLA violations but does not completely eliminate them.
\name{} yields the lowest cost -- 15\% less cost than autoscaling/Spock and cuts the cost by more than half compared to serverless-only (FaaS) or over-provisioning (MAX). 
In our simulation,
we assume that a serverless function has resources configured correctly so that each request always meets the SLA \cite{COSE}.
Thus, a serverless-only deployment yields zero SLA violations albeit at a higher cost. 
Similarly, \name{} yields zero SLA violations.
However, \name{} reduces the overall cost by always directing a portion of the demand to VMs that are provisioned to meet the SLA,
while the rest of the demand is directed to serverless functions that are also configured to meet the SLA.

\subsubsection{VM Provisioning \& Request Distribution}
\name{}'s main goal is to cautiously provision resources in the VM cloud (IaaS) to avoid under/over-utilization while simultaneously serving low-rate and sudden spikes in demand using serverless functions (FaaS).  
Figure~\ref{fig:vm_prov} shows how \name{} accurately tracks the incoming load, provisions VM resources, and avoids over-provisioning. 
This is observed by the similar behavior of \name{} in terms of the number of VMs provisioned (green solid curve) throughout the duration of the simulation and the {\em ideal} (offline) case (blue dashed curve) of provisioning the number of VM instances assuming perfect knowledge of future demand. 
The points on both curves represent scaling decisions taken every $K=300$ seconds (cf.\ Section~\ref{parameters}).

Figure \ref{fig:req_dist} shows the average rate of requests for the portion of the demand forwarded to the VM instances (IaaS) and the rest of the demand directed to serverless functions (FaaS) every 300~seconds. We observe that a consistent majority of the load is served by VMs (IaaS) whereas a small amount of the load with temporary peaks is handled by serverless (FaaS).

\subsubsection{VM Uptime \& Cost Breakdown}

\begin{figure}[h!]
\vspace{-3ex}
\hspace{-1.2em}
\captionsetup{justification=centering}
 \centering
  \subfloat[VM uptime]
  {\includegraphics[width=1.7in,height=1.3in]{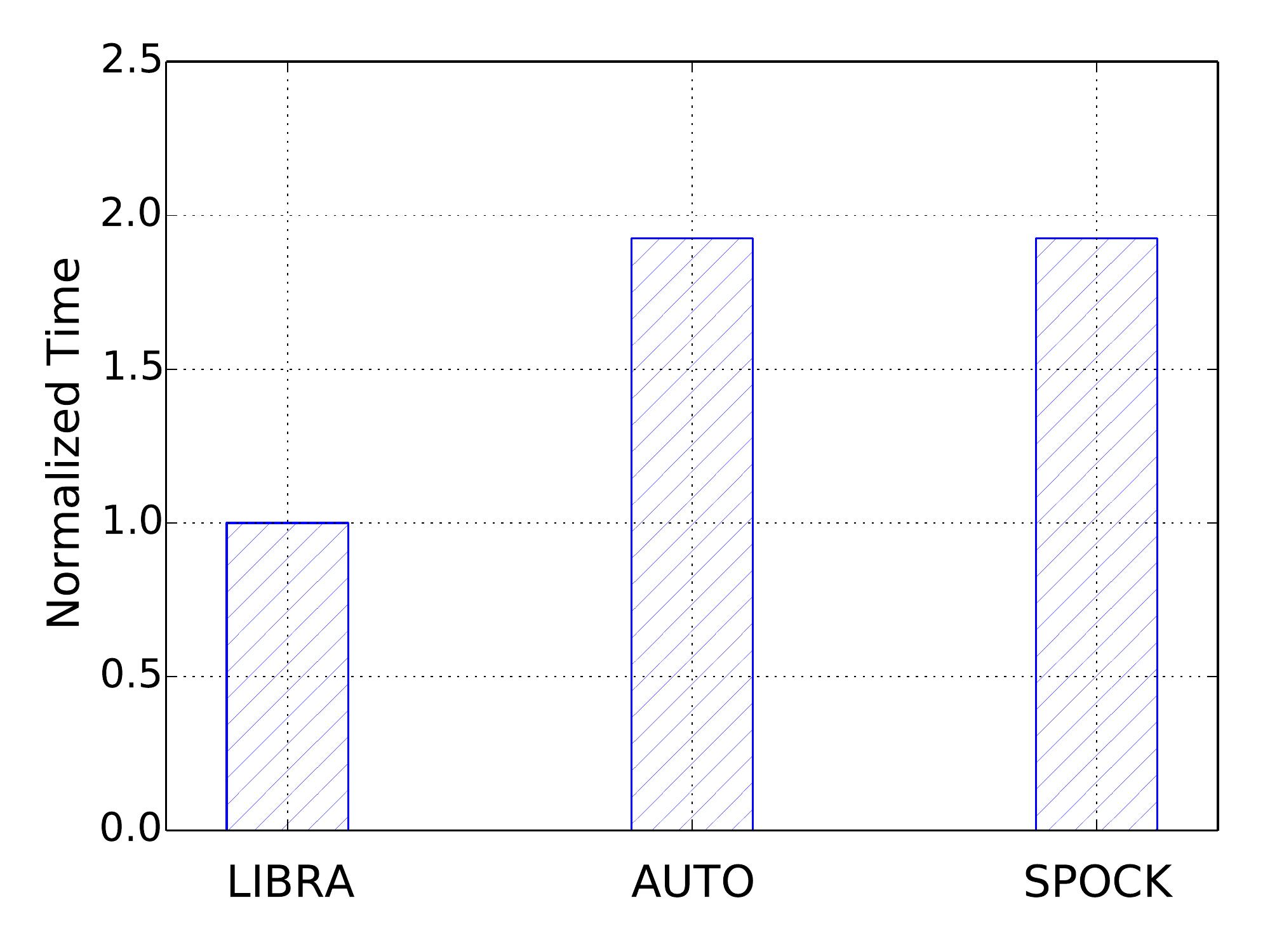}\label{fig:vm_uptime} \hspace{-0.6em}}
  \subfloat[Cost breakdown]
  {\includegraphics[width=1.7in,height=1.3in]{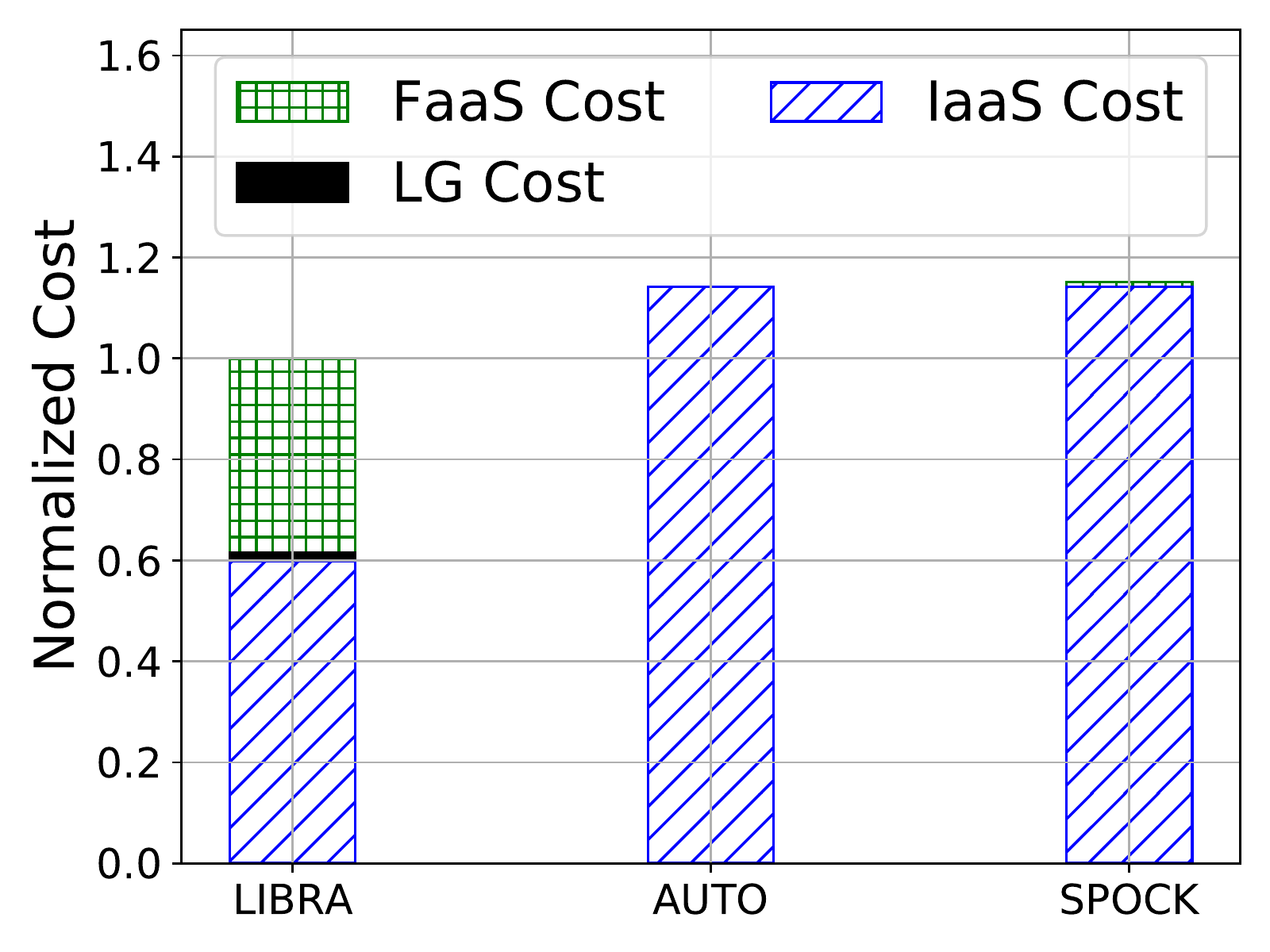}\label{fig:cost-breakdown} \hspace{-1.2em}  }
\caption{VM usage and cost breakdown} \label{fig:vm-and-cost}
\end{figure}

In Figure \ref{fig:vm_uptime}, we compare the total uptime of all VMs used to run the application under various approaches. \name{} cuts the VM uptime by half compared to autoscaling and Spock. 
This is because (1) \name{} only scales out when the demand persists for longer time. \name{} is able to identify transient demand and avoid reacting to it by using serverless functions (FaaS) rather than adding new VM instances (IaaS);
and (2) \name{} can add any arbitrary number of VMs to the active VMs when scaling out, while autoscaling adds or removes a user-configured number of instances (referred to as ``scaling group"). 
Figure \ref{fig:cost-breakdown} compares the 
cost breakdown across autoscaling, Spock, and \name{}. 
Autoscaling has zero FaaS cost since VMs are the only resources used to serve the application demand.
Spock employs FaaS when scaling out, only to hide VM startup delay, so the FaaS usage is really small ($\approx 1\%$). 
\name{} consistently uses serverless functions to serve a portion of the demand. The FaaS cost contributes around 40\% of the total cost in \name's case. 
Despite higher FaaS cost, the overall cost of \name{}
is smallest. 
\name{} intelligently uses FaaS for a portion of the demand that is either below CIP or transient,
since the cost of new VM instances for that portion of the demand would have been higher.
Note that the cost of \name{} includes that of the \name{} Gateway (LG), the added cost of running the \name{} system.

\subsubsection{Traffic Estimation}

\begin{figure}[h!]
\vspace{-2ex}
\captionsetup{justification=centering}
 \centering

  \subfloat[Total cost]
  {\includegraphics[width=1.6in, height=1.30in] {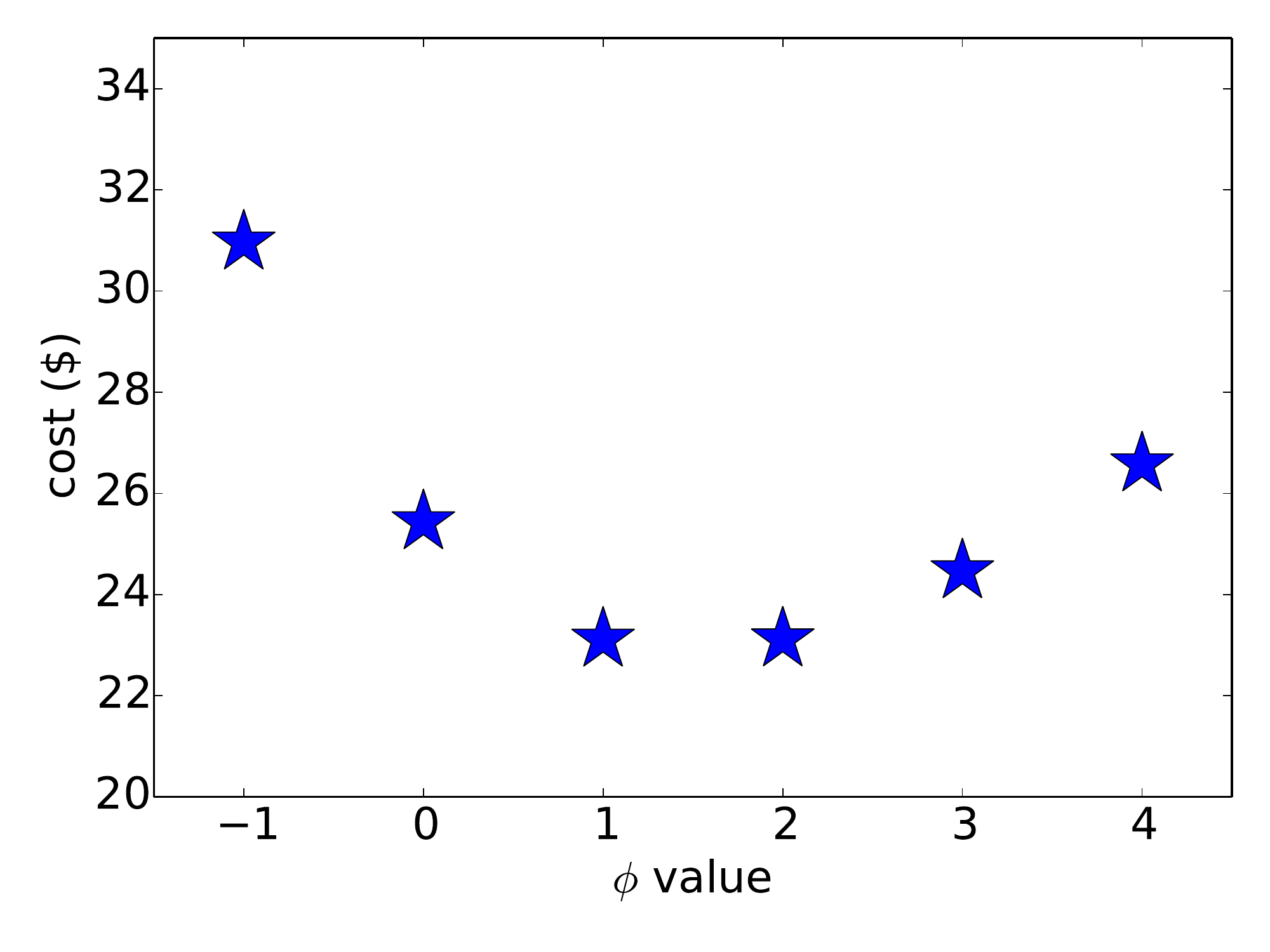}\label{fig:real_cost_perf} \hspace{-0.5em}} 
  \subfloat[Cost breakdown]
  {\includegraphics[width=1.6in, height=1.30in] {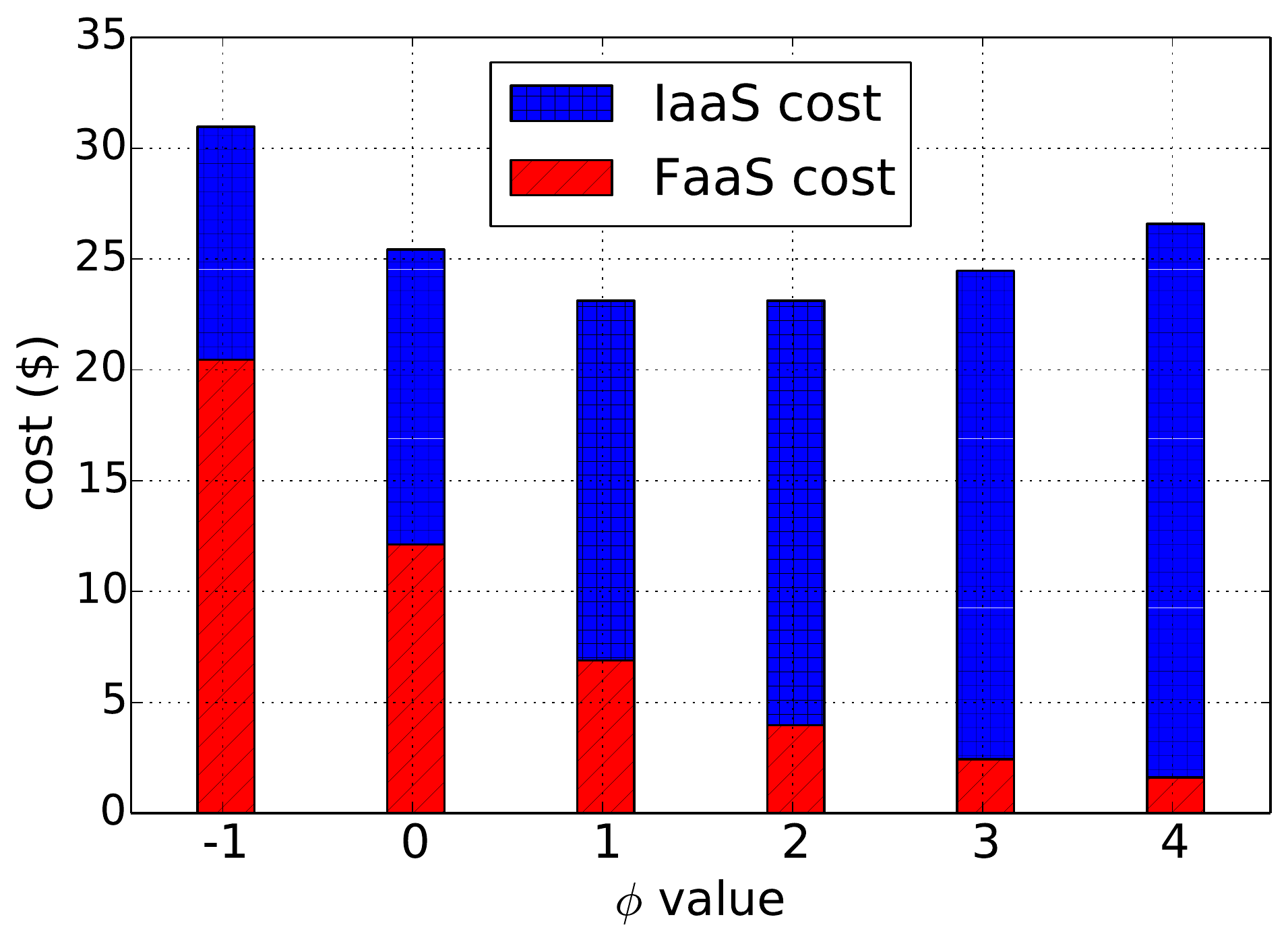}\label{fig:real_cdf}}
\caption{Cost for different values of $\phi$} \label{fig:rho}
\end{figure}

Recall that \name{} provisions IaaS resources for a request rate of $vm_{reqs} = (avg+ \phi \cdot std)$.
The actual request rate directed to the provisioned VMs is $\rho \cdot vm_{reqs}$ ($\rho < 1$),
while the remaining requests (in each epoch) are directed to FaaS where they are served within the SLA but at a higher cost. 
The value of $\phi$ can be adapted based on the particular fluctuations in demand.
Tuning $\phi$ affects the cost but not
the performance of an application.
This is because the SLA is met whether 
\name{} directs the request to the {\em provisioned} VMs or to FaaS,
however FaaS is more costly. For the WITS traces used in our evaluation, the effect of different values of $\phi$ on the cost is shown in Figure~\ref{fig:rho}. We observe that a lower value of $\phi$ leads to more FaaS usage and hence higher overall cost, whereas a higher value of $\phi$ causes over-provisioning of VMs, which leads to VM under-utilization and hence higher cost. 
Here, $\phi = 1$ gives the least cost.

\section{\name{} on AWS}
\label{sec:aws}

To validate our simulation results from Section~\ref{sec:simulation},
where \name{} was shown to be effective in reducing both cloud-usage cost and SLA violations, we implemented \name{} to perform load balancing for an application deployed on Amazon AWS Cloud, {\em i.e.}\ EC2 for IaaS and Lambda for FaaS.

\subsection{Application}
The application is an image manipulation application written in Python. On Amazon Lambda, we are able to deploy the application as a single function. While we expect similar results for multi-function applications, our choice of single-function application is inspired by many use cases such as ML inference models~\cite{spock, mark, ishakian2018serving}, IoT and computer vision applications~\cite{lambda_rekognition} which can be usually deployed as a single function. 
To deploy the application on EC2 VM instances, we used a python multi-threaded HTTP server library. 
We ran it on a {\em t3.medium} \cite{EC2_instance_type} EC2 instance type with Ubuntu Server 18.04 LTS operating system, two 2.5 GHz vCPU, and 4 GB memory. 

\subsection{Application Profiling \& Lambda Resources}
As described in Section \ref{scaling-manager}, \name{}'s {\em Scaling Manager} uses the maximum number of requests, $r_{max}$, that a VM instance can handle, to calculate the required number of instances. To obtain $r_{max}$ for this evaluation, we deployed our application on a {\em t3.medium} instance, profiled its performance, and obtained $r_{max}$ that meets the SLA. We take the SLA to be one second of execution time serving a request. Profiling an application on a given VM instance is a one-time task and a developer can perform this prior to production deployment. For serverless functions, we configured the memory such that it, which is the only configurable resource for AWS Lambda. We invoked the function with various memory configurations and picked the memory setting that gave the least cost while meeting the SLA.

\subsection{Setup \& Implementation}
To obtain a consistent network environment for our evaluation on AWS, we deployed an application client on an EC2 instance, which will generate HTTP requests for the application. We compare \name{} to the same four resource provisioning and deployment strategies described in Section~\ref{sec:policies}.
We needed to modify the implementations of  \name{}, Autoscaling, and Spock, as follows:
\subsubsection{LIBRA on AWS}
We deployed the \name{} Gateway (LG) on an EC2 instance of the type {\em t2.micro}.
The LG distributes requests between IaaS based resources (EC2 VMs) and FaaS (AWS Lambda functions). 
Within IaaS, we used the AWS Application Load Balancer (ALB) \cite{aws_alb} to distribute requests among the active VM instances evenly, {\em i.e.}\ in a Round Robin fashion.
\subsubsection{Autoscaling on AWS}
 We used EC2's autoscaling service, which is a threshold-based scaling service. The AWS CloudWatch Alarm was used to monitor {\em RequestCountPerTarget} metric to make scaling decisions. Again, ALB was used to distribute requests to the VMs.
\subsubsection{Spock on AWS}
An alarm from AWS CloudWatch triggers the scale-out or scale-in events. 
When a scale-out event is triggered, new VM instances are provisioned. During this VM provisioning time, the system sends the extra requests, that cannot be served by the active VM instances, to the serverless functions. 
Once the new VM instances are ready, all requests are forwarded to the VMs. Thus, Spock attempts to use serverless functions (FaaS) only to hide VM startup delays.

\subsection{Results \& Discussion}
For our AWS experiments, to reduce real load,
we use a scaled-down version of
the first 1800 seconds of the WITS trace 
shown in Figure \ref{fig:traces}. 
In particular, we reduce the rate of requests by a factor of 16. 
\begin{figure}[h!]
\vspace{-2ex}
\captionsetup{justification=centering}
 \centering
 \subfloat[Cost and SLA violations for \name~vs. other policies]
  {\includegraphics[width=1.8in, height=1.36in] {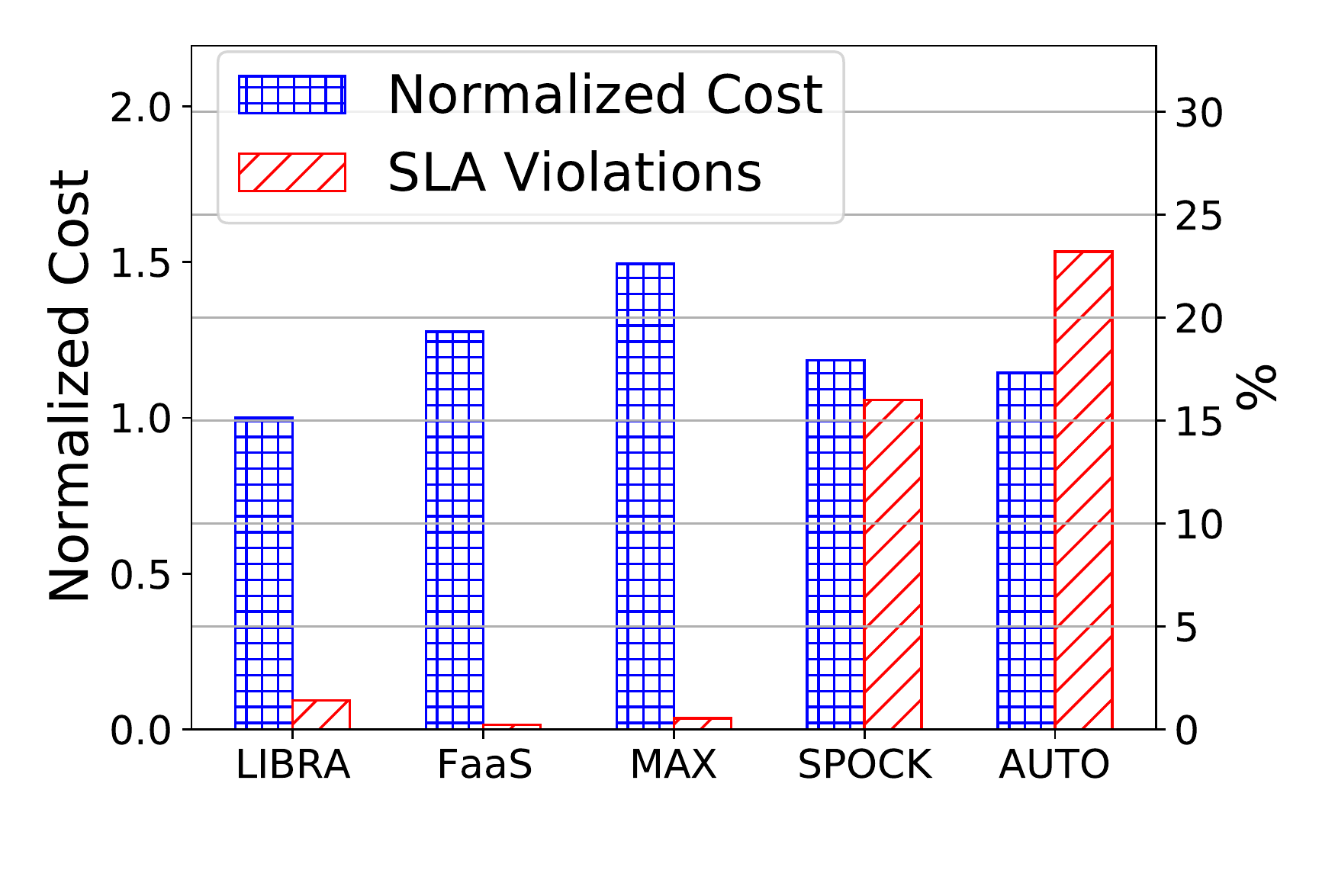}\label{fig:real_cost_perf} \hspace{-0.8em}} 
  \subfloat[Response time distribution]
  {\includegraphics[width=1.8in, height=1.4in] {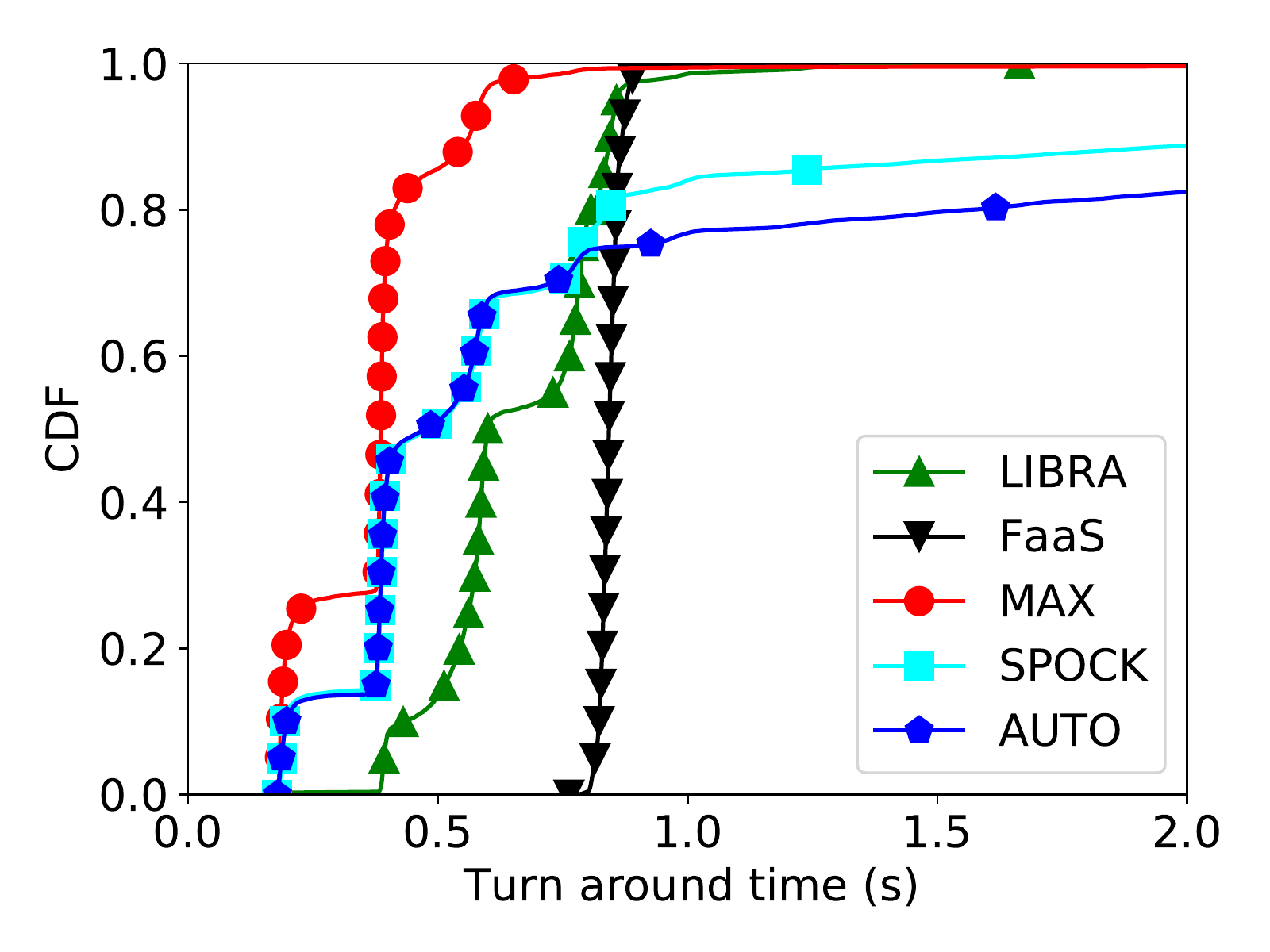}\label{fig:real_cdf}}
\caption{Performance of \name{} on AWS} \label{fig:perf_all}
\end{figure}

\subsubsection{Cost and SLA Violations} 
We compare the cost and performance of \name{} versus other resource provisioning and deployment strategies. The results are consistent with our simulation results. Figure \ref{fig:real_cost_perf} shows that \name{} yields the lowest cost with very low amount of SLA violations. \name{} reduces the SLA violations (by more than 85\%) and cost (up to 20\%) when compared to auto-scaling and Spock. \name's{} cost also includes the cost of deploying the \name{} Gateway on an EC2 instance. Max-provisioning and serverless-only deployment yield the lowest SLA violations but incur up to 50\% increase in cost. 
We observe that FaaS, \name{}, and max-provisioning, all have a little amount of SLA violations. 
This is because unlike our simulation model, in a real setup, 
factors such as co-location, 
cold-starts for serverless functions, 
and underlying resource contention for VMs, can introduce slight variation in the performance of an application. 
For \name{}, a lower value for the VM utilization parameter $\rho$ (discussed in Section~\ref{vm-util}) can mitigate these SLA violations.

While Spock reduces SLA violations by 40\% compared to autoscaling, about 15\% of the requests fail to complete within the SLA. This can be explained by Spock's reactive scheme, where scaling out is triggered when VM resources are saturated, resulting in SLA violations.
On the other hand, \name{} avoids saturating the VM instances by directing excess load to serverless functions. 
If the load is not transient and the demand stays higher for a longer period, \name{} increases the number of VM instances at the next scaling decision.

Figure \ref{fig:real_cdf} shows the CDF of completion time of each request. The over-provisioning policy gives the best performance in terms of keeping execution time really low, as expected. FaaS has the most consistent performance, {\em i.e.}\ the completion time is always between 0.8s to 1s. This is due to the fact that each request is executed in a dedicated sandbox environment with dedicated resources, such as memory, so the chance of performance fluctuation is lower. The performance of a serverless function is primarily affected by cold-start delay \cite{ishakian2018serving},
which is negligible and can be as low as 10s of milliseconds for serverless functions written in Python \cite{peeking-behind}.

\begin{figure}[h!]
\vspace{-2ex}
\hspace{-1.2em}
\captionsetup{justification=centering}
 \centering
  \subfloat[VM uptime]
  {\includegraphics[width=1.7in,height=1.2in]{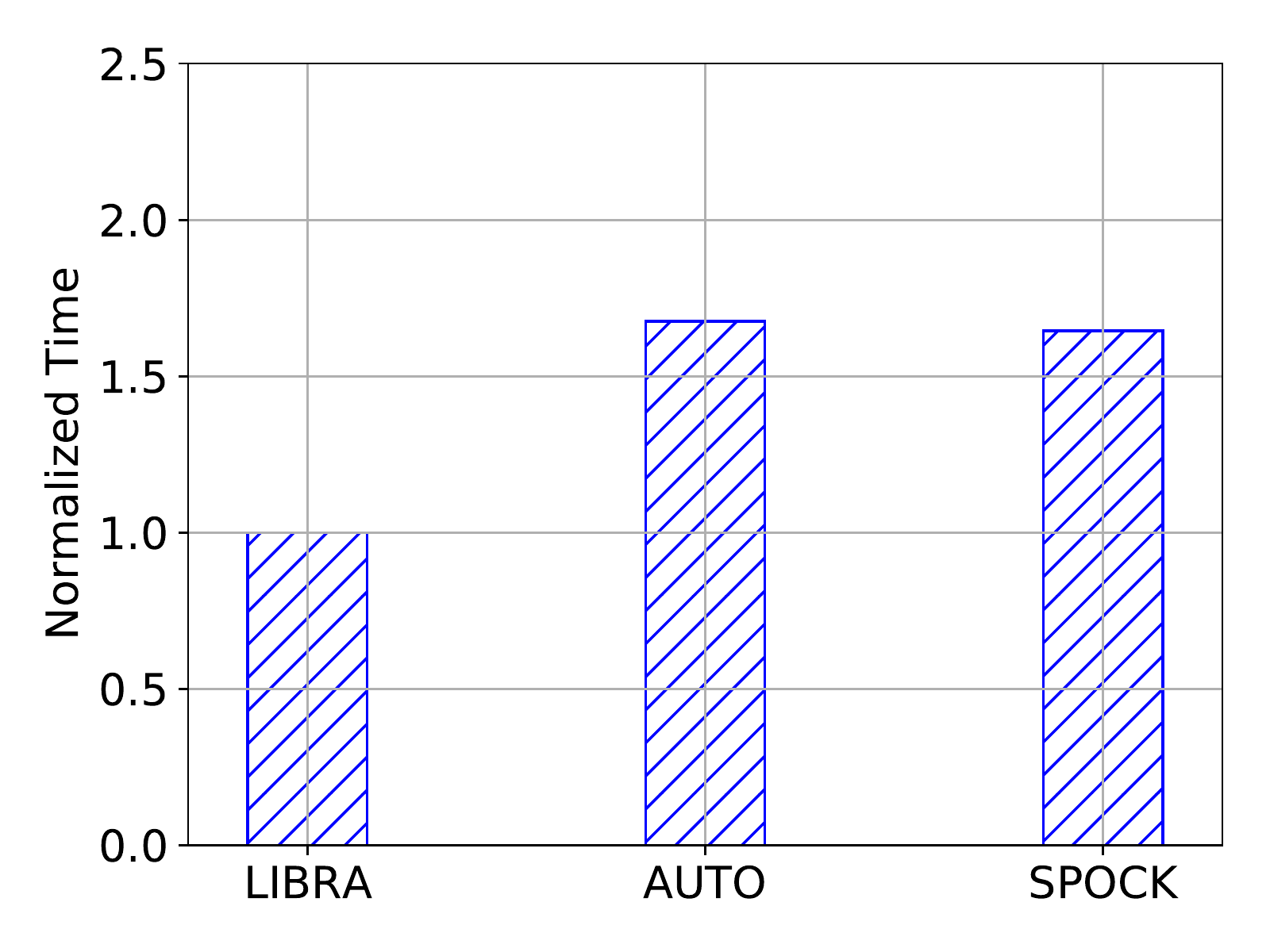}\label{fig:real_vm_uptime} \hspace{-0.6em}}
  \subfloat[Cost breakdown]
  {\includegraphics[width=1.7in,height=1.2in]{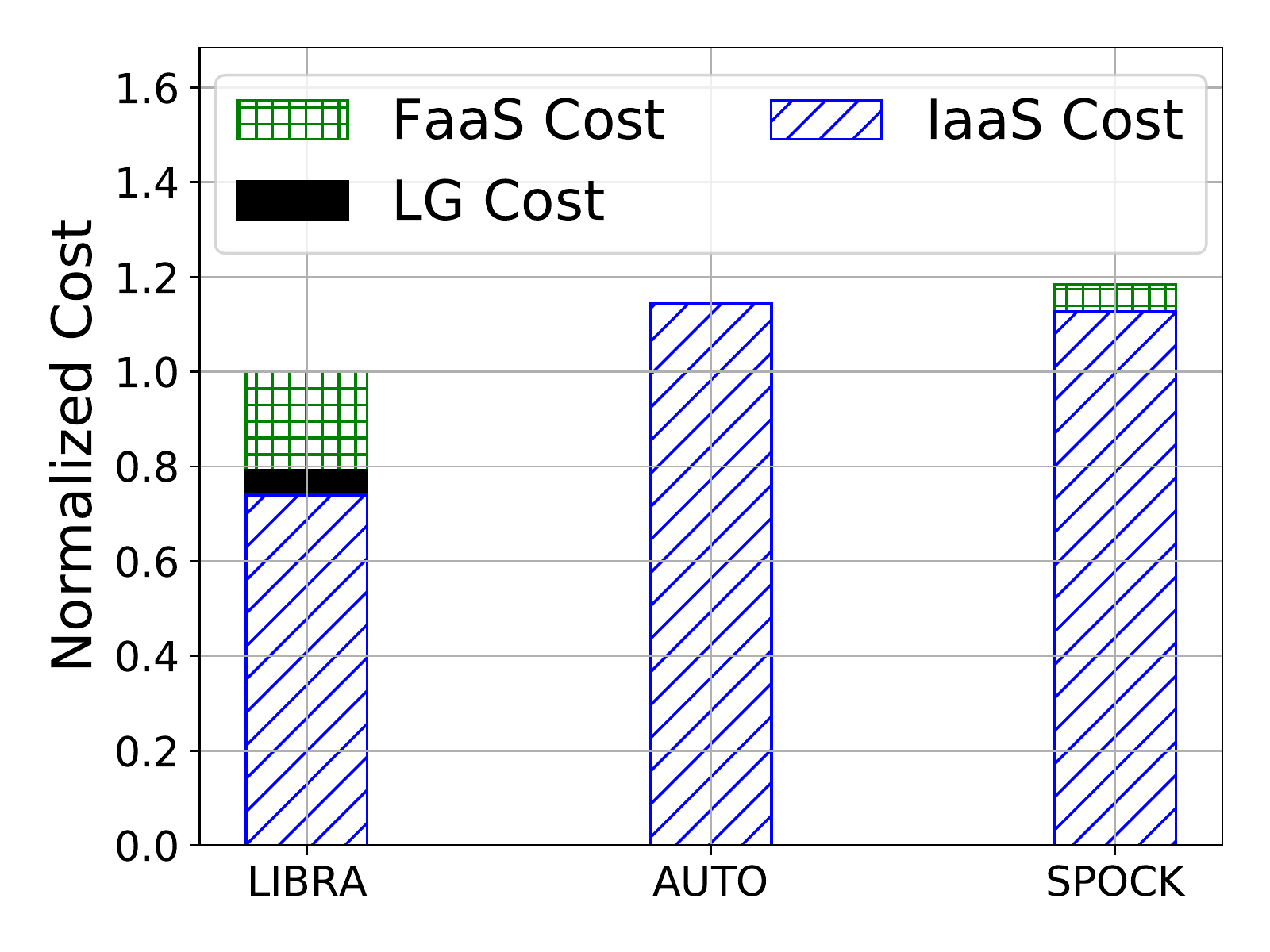}\label{fig:real_cost_breakdown} \hspace{-1.2em}  }
\caption{VM usage and cost breakdown} \label{fig:break-even-result}
\end{figure}

\subsubsection{VM Uptime \& Cost Breakdown}
Figure \ref{fig:break-even-result} confirms the benefit of \name{} by illustrating the cost breakdown. This is consistent with our simulation results. Figure \ref{fig:real_vm_uptime} shows the uptime of VM instances used to run the application for Spock, Autoscaling, and \name{}. 
\name{} is able to closely monitor the demand and provision required VM resources without over-provisioning, which results in lower overall VM uptimes and cost. Figure \ref{fig:real_cost_breakdown} shows the cost breakdown of these approaches. \name{} has the lowest overall cost (including the cost to deploy LG), with lowest IaaS cost but highest FaaS cost. 
\name{} uses serverless functions more consistently and effectively, {\em i.e.}\ for transient demand or portion below CIP,
which results in higher usage of FaaS, and 
lower usage of IaaS.

\subsection{Performance Overhead \& \name~Scalability}
\name{} works as a legacy load balancer and directs requests to appropriate resources, 
introducing overhead no more than legacy load balancers. 
At the same time, other \name{} operations, {\em i.e.} scaling and traffic monitoring,
occur in the background and do not impact the real-time processing of requests.
The \name{} Gateway has a small computational footprint, 
and hence a small cost. 
Various components of the \name{} Gateway can be implemented as serverless functions, 
where scalability is taken care of by the cloud provider. 
Alternatively,
the \name{} Gateway can be implemented as one service and deployed using VM instances and the application administrator can rely on the cloud provider's scaling services such as AWS auto-scaling.

\section{Related Work}
\label{sec:related-work}
Recent studies have shown that serverless is a reliable alternative to develop and deploy cloud applications, hence establishing the practicality of \name{}. MArk\cite{mark}, Spock\cite{spock}, and Costless\cite{costless} show various ML applications/models can be deployed using serverless functions. Other approaches explore the use of serverless functions in building video/data processing \cite{video_processing_ao,video_processing_zhang, data_processing_jonas} and IoT applications~\cite{serverless-iot-1}. 
Malawski et al. \cite{sci-workflow} show that serverless functions can be used to implement scientific workflows. Other studies leverage the on-demand computation and scalability of serverless functions for biomedical applications~\cite{dna-ser-1, dna-ser-2, dna-ser-3} and to solve various mathematical and optimization problems~\cite{regression-in-ser, pywren, ser-big-data}.

CherryPick \cite{cherry-pick} helps the developer find the best VM instance type using statistical learning techniques. Previous approaches use various ML and learning-based approaches to configure serverless functions \cite{COSE, ai_con_per, sizeless}. Costless \cite{costless} explores different function placement solutions to deploy a data fusion application on AWS Lambda to minimize cost and meet performance constraints. CloudCmp \cite{cloud-cmp} performs comprehensive measurement studies on various commercial clouds so as to find a suitable cloud provider for a given application. Similarly, cluster management systems, such as Google Borg \cite{borg} and Mesos \cite{mesos}, orchestrate and allocate resources in the cloud for various applications. Google has developed a machine type recommendation system \cite{google_cloud_rec} that helps a user find the suitable instance type that maximizes resource utilization. Moreover, most of the commercial cloud providers provide autoscaling services (reactive in nature) to manage the virtual resources, {\em e.g.,}\ AWS's EC2 autoscaling \cite{amzone_auto_scaling} and Google autoscaling \cite{google_auto_scaling}. Moreover, similar to \name{}, other approaches use various prediction models to predict the demand for an application and orchestrate cloud resources\cite{rpps, predictive-models, myse, svm-prediction,bats}. The aforementioned approaches only manage resources within one type of service , while \name{} orchestrates resources across two different services, {\em i.e.},\ an IaaS and a FaaS.

Previous works have used serverless functions to hide VM startup delays for ML or other applications, while scaling out VM-based resources~\cite{spock,mark,feat}.  In contrast, \name{} uses serverless functions as an alternative cloud service to run applications, and based on the demand, can decide to use either one or both services to optimize {\em both} cost and performance.

\section{Conclusion and Future Work}
\label{sec:conc}

We presented \name{}, a load balancing approach that simultaneously uses both IaaS and FaaS cloud services to cater to the dynamic demand of applications. Our approach is application agnostic and can be employed by cloud providers as a value-added service or used by  end-users directly. Our evaluation of \name{} in simulations and on AWS shows its clear advantage over other resource provisioning policies in reducing both cost of cloud usage and SLA violations.

In this paper, \name{} utilizes IaaS and FaaS services available from one cloud provider in support of single-function applications.
Future work includes extending \name{} to accommodate applications that consist of chains of functions,
and to utilize services from multiple cloud providers.
\bibliographystyle{IEEEtran}
\bibliography{main}
\end{document}